\documentclass[%
 aip,
 reprint,
 amsmath,amssymb,
]{revtex4-2}

\usepackage[T1]{fontenc}
\usepackage{times}
\usepackage{graphicx}
\usepackage{bm} 
\usepackage{bbm}
\usepackage{amsfonts, amsmath,amssymb, bm, graphicx}
\usepackage{siunitx}
\usepackage{ulem}
\usepackage{amssymb}
\usepackage[english]{babel}
\usepackage{lipsum}
\usepackage{mathtools}
\usepackage{dcolumn} 
\usepackage{bm} 
\usepackage[english]{babel}
\usepackage{amssymb,amsmath}
\usepackage{hyperref}
\usepackage{color}

\usepackage{soul}

\begin{document}
\preprint{AIP/123-QED}

\title{Exchange-dominated frequency shift of spin-wave nonreciprocal dispersion relation in planar magnetic multilayers}

\author{Claudia Negrete}
\affiliation{Departamento de F\'isica, Universidad Cat\'olica del Norte, Avenida Angamos 0610, Antofagasta, Chile}

\author{Attila K\'{a}kay}
\affiliation{Helmholtz-Zentrum Dresden~-~Rossendorf, Institute of Ion Beam Physics and Materials Research, Bautzner Landstr. 400, 01328 Dresden, Germany}

\author{Jorge A. Ot\'alora}
\email[Corresponding author: ]{jorge.otalora@ucn.cl}
\affiliation{Departamento de F\'isica, Universidad Cat\'olica del Norte, Avenida Angamos 0610, Antofagasta, Chile}

\date{\today}

\begin{abstract}
Spin-wave nonreciprocity, manifested as a frequency difference between counterpropagating modes, underpins many proposed magnonic devices. While this effect is commonly attributed to dipolar interactions or interfacial chirality, the microscopic origin of the frequency shift in nonreciprocal dispersion in magnetic multilayers remains under debate. Here, we analyze the frequency shift of nonreciprocal spin-wave dispersion in planar multilayer heterostructures without Dzyaloshinskii-Moriya interaction. Using a frequency-shift dynamic matrix  formalism, we show that the frequency asymmetry cannot generally be ascribed to dipolar effects alone. Instead, once counterpropagating modes differ in their geometric structure along the thickness, interlayer exchange dominates the frequency shift. Applied to representative multilayer systems, we find that the interlayer exchange contribution exceeds dipolar and intralayer exchange effects by up to two to three orders of magnitude over a broad wave-vector range. Our results establish interlayer exchange as the primary mechanism contributing to the frequency shift of nonreciprocal dispersion in multilayer magnonic systems and provide a quantitative framework for engineering large frequency shifts in nonreciprocal magnonic devices.
\end{abstract}

\maketitle

\section{Introduction}
In the context of spin waves (SWs) - the collective eigenexcitations of an arrangement of magnetic moments - nonreciprocity is usually understood as the frequency difference between counterpropagating SWs with the same wave-vector magnitude $k$. In such cases, the dispersion relation (i.e., the SW oscillation frequency $\omega$ as a function of the wave vector $\mathbf{k}$) is said to be nonreciprocal, $\omega(\mathbf{k})\neq\omega(-\mathbf{k})$, leading to a finite frequency difference $\Delta\omega(\mathbf{k})=\omega(\mathbf{k})-\omega(-\mathbf{k})\neq 0$. Identifying and controlling the conditions that give rise to this nonreciprocal effect has been the subject of extensive theoretical and experimental efforts \cite{JOURChumak,APLWang2014,Chen_2022,CAMLEY1987103,SAVCHENKO20199}. This interest is largely motivated by the fact that nonreciprocity constitutes a fundamental mechanism for enhancing the capabilities of sensing, processing, transmission, and buffering of information encoded in the oscillatory properties of SWs in magnetic systems.\cite{Chumak2022IEEE}
Examples of applications include magnetic circulators and isolators for directional signal control \cite{artFusheng,Wang2021,109Zenbaa}, the suppression of backscattering noise in SW logic gates and magnetic computing circuits \cite{Jamali2013,Khitun2010,105Khitun7}, nonreciprocal passive filters that improve signal quality by eliminating interference in a specific propagation direction in wireless networks \cite{arHarward2014,artFranco2020}, and spin-wave-based diodes that enable unidirectional wave transmission, which is essential for controlled signal routing in magnetic circuits for wave-based computing \cite{GrassiDIODO2020,ZouDiode2024}.

Over the past decade, significant efforts have been devoted to engineering dispersion nonreciprocity, including  curving magnetic membranes \cite{Korber2022,otalora2017,Salazar2021,PhysRevApplied.18.054044}, grading the saturation magnetization in planar magnetic films \cite{Sluka2018EmissionAP,Gallardo2019}, stacking membranes composed of different magnetic materials \cite{Christienne25AnisotropyGraded,GrassiDIODO2020}, and piling up magnetic membranes exhibiting intrinsic spin-orbit interactions, such as the Dzyaloshinskii-Moriya interaction (DMI) \cite{MimicaDipoleDM25,Zakeri2010,Kuepferling2023}, among others. Despite this progress, a striking asymmetry exists in how different magnetic interactions are treated in the interpretation of nonreciprocal dispersion. In most studies, dipolar interactions-and, where present, DMI are identified as the primary origin of frequency nonreciprocity, while exchange and anisotropy interactions are assumed to play at most a secondary or indirect role ~\cite{PhysRevB111134434,PhysRevApplied21014035}. This interpretation is often motivated by the observation that only dipolar or chiral interactions generate explicitly wavevector-odd contributions to the dynamic matrix. However, such reasoning implicitly assumes that counterpropagating SW modes are geometrically equivalent, an assumption that does not generally hold in multilayer or vertically inhomogeneous systems.  

Recently, symmetry-based approaches using quantities such as the toroidal moment have been proposed to predict dispersion nonreciprocity based solely on the equilibrium magnetic configuration and the propagation direction~\cite{Cheong2018,Cheong2022}. While powerful, these approaches do not explicitly resolve how individual interactions contribute to the frequency shift once the full dynamical mode structure is taken into account. In particular, the role of exchange interactions in nonreciprocal systems - especially interlayer exchange in multilayer stacks - remains insufficiently understood.  

In this work, we address this gap by analyzing the full interaction-resolved origin of spin-wave frequency shifts in planar heterostructured multilayer magnetic systems without DMI. Using a dynamic-matrix formalism for Damon-Eshbach spin waves, we demonstrate that exchange interactions can dominate the nonreciprocal frequency shift whenever counterpropagating modes differ in their geometric structure along the thickness. By introducing a frequency-shift dynamic matrix, we show that the observed nonreciprocity results from an exchange-dominated energetic imbalance driven by dipolar-induced mode asymmetry. This provides a unified physical mechanism that reconciles seemingly dipolar-driven dispersion asymmetries with an underlying exchange origin, and applies broadly to experimentally relevant multilayer systems.

\smallskip{}

\begin{figure}
\includegraphics[scale=0.48]{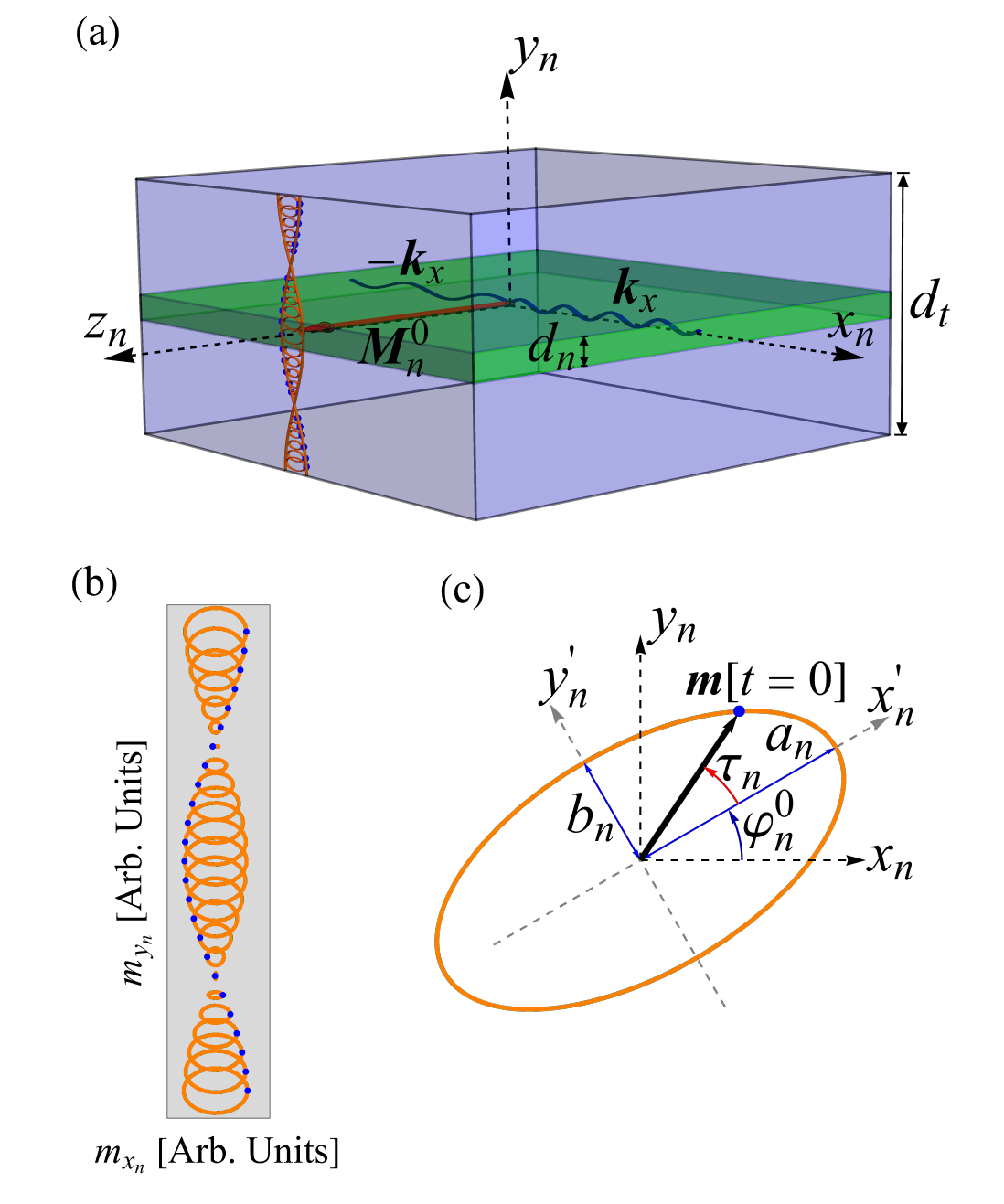}
\caption{Illustration of (a)  a homogeneous magnetization state with a total thickness $d_{t}$, (b) spin-wave orbit distribution along the thickness of the third excitation mode, and (c) elliptical spin wave orbit at the n'th sublayer. }
\label{fig:system}
\end{figure}

\section{Theoretical model}
The system under consideration is described within the continuum micromagnetic framework using the dynamic matrix method (DMM). We consider a planar heterostructure composed of a multilayer stack of $N$ magnetically coupled sublayers. The magnetization dynamics of each sublayer is governed by the Landau-Lifshitz (LL) equation, resulting in a system of $N$ coupled LL equations. The $\mathrm{n'th}$ sublayer is characterized by a magnetization vector $\mathbf{M}_\mathrm{n}$, an exchange length $l_{\mathrm{ex}_n}$, an exchange stiffness constant $A_{n}$ and a saturation magnetization $M_{s_{n}}$. 

Figure~\ref{fig:system}(a) illustrates the geometry of the system, which has a total thickness $d_t$. The $n$th sublayer has thickness $d_n$ and lies in the $y_n z_n$ plane. Counterpropagating spin waves propagate along the $\hat{x}_n$ direction with wave-vector magnitude $k_x$. The following assumptions are made throughout this work:
\begin{enumerate}
\item an external magnetic field $\mathbf{H}=\mathrm{H}\,\hat{z}_n$ is applied;
\item an in-plane uniaxial anisotropy characterized by the constant $K_{n}^{u}$ favors magnetization alignment along the $\hat{z}_n$ axis;
\item the isotropic exchange interaction is decomposed into an \textit{intralayer exchange} acting within a single layer in the discretization scheme, characterized by the exchange length $l_{\mathrm{ex}_n}$,
\item and into an \textit{interlayer exchange}, responsible for the coupling between the $n$th and $\beta$th sublayers, described by the integral exchange constant $J_{n,\beta}$;
\item the equilibrium magnetization state is a uniform monodomain state lying in-plane along the easy axis $\hat{z}_n$, given by $\mathbf{M}_n^0=M_{s_n}\hat{z}_n$, as shown in Fig.~\ref{fig:system}(a)
\end{enumerate}

We emphasize that the interlayer exchange interaction defined here should not be confused with the Ruderman-Kittel-Kasuya-Yosida (RKKY)-type interlayer exchange coupling. In the present work, the interlayer exchange refers exclusively to the conventional isotropic exchange interaction acting along the direction perpendicular to the film plane.

The LL equation governing the dynamics of the $\mathrm{n'th}$ sublayer reads
\smallskip{}
\begin{equation}
    \dot{\mathbf M}_n[\mathbf r,t]=-\mu_0\gamma\mathbf{M}_n[\mathbf r, t]\times\mathbf{H}_n^e[\mathbf r,t],
    \label{eq:LLG}
\end{equation}
\smallskip{}
\noindent where  $\gamma$ is the magnitude of the gyromagnetic ratio, $\mathbf{H}_{n}^{e}[\mathbf{r},\,t]$  is the effective magnetic field acting on the $n$-th layer, and $\mathbf{M}_{n}[\mathbf{r},\,t]$ is the magnetization field. 
The spin-wave dispersion relation and the equilibrium homogeneous state are obtained by linearizing the LL-equation around the equilibrium magnetization $\mathbf{M}_n^0$. Accordingly, the magnetization and the effective field are expanded as $\mathbf{M}_{n}[\mathbf{r},\,t]=\mathbf{M}_{n}^{0}+\mathbf{m}_{n}[\mathbf{r},\,t]$ and $\mathbf{H}_{n}^{e}[\mathbf{r},\,t]= \mathbf{H}_{n}^{0e}+\mathbf{h}_{n}^{e}[\mathbf{r},\,t]$, respectively. Here, $\mathbf{H}_{n}^{0e}$ denotes the equilibrium effective magnetic field, while $\mathbf{m}_{n}[\mathbf{r},\,t]$ and $\mathbf{h}_{n}^{e}[\mathbf{r},\,t]$ are the dynamical perturbations. We assume plane-wave solutions of the form $\mathbf{m}_{n}[\mathbf{r},t] =  \int d^3\mathbf{k}\,e^{i(\omega t-\mathbf{k}\cdot \mathbf{r})} \tilde{\mathbf{m}}_n[\mathbf{k}]$, and $\mathbf{h}_{n}[\mathbf{r},t] =  \int d^3\mathbf{k}\,e^{i(\omega t-\mathbf{k}\cdot \mathbf{r})} \mathbf{h}_n[\mathbf{k}]$, where $\tilde{\mathbf{m}}_n[\mathbf{k}]$ and $\mathbf{h}_n[\mathbf{k}]$ are the dynamic magnetization and magnetic field in wave vector $\mathbf{k}$ space, respectively, and $\omega$ is the angular resonance frequency of the spin waves. 

Linearization of the LL equation around the equilibrium state $\mathbf{M}_n^0$ yields the eigenvalue problem 
\smallskip{}
\begin{equation}
\omega\,\tilde{\mathbbm{m}}[\mathbf{k}] = \mathbb{N}[\mathbf{k}] ~\tilde{\mathbbm{m}}[\mathbf{k}]\label{eq:LINEALIZA_LL},
\end{equation}

\noindent where $\mathbb{N}$ is the dynamic matrix of dimension $2N\,\times\,2N$, with $N$ the number of sublayers, $\omega$ is the eigenvalue, and  $\tilde{\mathbbm{m}}[\mathbf{k}]$ is the corresponding eigenvector with transpose $\tilde{\mathbbm{m}}[\mathbf{k}]^{T}=\left\{ \tilde m_{\text{X}_{1}}[\mathbf{k}],\text{...},\tilde m_{\text{X}_{N}}[\mathbf{k}],\tilde m_{\text{Y}_{1}}[\mathbf{k}],\text{...},\tilde m_{\text{Y}_{N}}[\mathbf{k}]\right\}$. A representative second-order spin-wave mode profile across the multilayer thickness is shown in Fig.~\ref{fig:system}(b), where the blue dot point indicates the dynamic magnetization at zero time ($\mathbf{m}_n[t=0]$). The SW orbit at the $n$-th layer is also illustrated in Fig.~\ref{fig:system}(c), where the geometrical parameter as large(smaller) semiaxis $a_n$($b_n$), tilting angle $\varphi_n^0$ and zero-time phase (also named here as phase) $\tau_n$ are shown as well. 

Explicit expressions for the dynamic matrix $\mathbb{N}$ in multilayer systems have been reported previously \cite{Gallardo2019,Christienne25AnisotropyGraded}. In the following, we therefore focus on the specific matrix components relevant to  Damon-Eschbach (DE) spin waves, for which the wavevector $\mathbf{k}=k_x\hat x$ is perpendicular to the equilibrium magnetization $\mathbf{M}_0=M_{s_n}\hat z$ (see Fig.~\ref{fig:system}(a)).

The total dynamic matrix can be expressed as the sum of contributions from dipolar (dip), intralayer exchange (ex), interlayer exchange (int), uniaxial anisotropy (U), and Zeeman (Ze) interactions,
\begin{equation}
\mathbb{N}=\mathbb{N}^{\mathrm{dip}}+\mathbb{N}^{\mathrm{ex}}+\mathbb{N}^{\mathrm{int}}+\mathbb{N}^{\mathrm{U}}+\mathbb{N}^{\mathrm{Ze}}
\end{equation}

Each contribution has a block structure
\smallskip{}
\begin{equation}
\mathbb{N}^\sigma =
\begin{pmatrix}
\mathbb{N}_{\mathrm{XX}}^{\sigma} & \mathbb{N}_{\mathrm{XY}}^{\sigma} \\
\mathbb{N}_{\mathrm{YX}}^{\sigma} & \mathbb{N}_{\mathrm{YY}}^{\sigma}
\end{pmatrix}
\end{equation}
\smallskip{}
\noindent with $\sigma\in\left\{\mathrm{dip,  int, ex, U, Ze}\right\}$. The dynamic matrix blocks $\mathbb{N}_{q}^\sigma$ are defined as
\smallskip{}
\begin{equation} \label{eq:matrixdinamica}
\mathbb{N}_{q}^{\sigma} =
\begin{pmatrix}
N_{11}^{q,\sigma} & N_{12}^{q,\sigma} & \cdots & N_{1N}^{q,\sigma} \\
N_{21}^{q,\sigma} & N_{22}^{q,\sigma} & \cdots & N_{2N}^{q,\sigma} \\
\vdots & \vdots & \ddots & \vdots \\
N_{N1}^{q,\sigma} & N_{N2}^{q,\sigma} & \cdots & N_{NN}^{q,\sigma}
\end{pmatrix},
\end{equation}
\smallskip{}
\noindent where $q\in\left\{\mathrm{XX,XY,YX,YY}\right\}$, and $N_{np}^{q,\sigma}$ with $n,p=1,2,\cdots N$  are the dynamic matrix entries of $\mathbb{N}_{q}^{\sigma}$. Solving Eq.~\eqref{eq:LINEALIZA_LL} yields both the dispersion relation $\omega[\mathbf k]$ and the spin-wave eigenvectors $\tilde{\mathbbm{m}}[\mathbf{k}]$.

The dynamic matrix entries of the individual interactions are shown in the supplementary material (SM Section III)\cite{suplementario}. However, for the completeness of the manuscript, are also summarized in the following. For the dipolar, intralayer exchange and uniaxial anisotropy interactions, the dynamic matrix entries take the form
\smallskip{}
 \begin{align}
    N_{np}^{\mathrm{XX}, \rm a} &= -\mathrm{i} \,  \omega_{s_n} A^{\mathrm{XY},\rm{a}}_{np},\label{eq:Nxxa}\\
    N_{np}^{\mathrm{XY}, \rm a} &= -\mathrm{i} \, \omega_{s_n}A^{\mathrm{YY},\rm{a}}_{np}, \label{eq:Nxya}\\
    N_{np}^{\mathrm{YX}, \rm a} &= \mathrm{i} \,\omega_{s_n}A^{\mathrm{XX},\rm{a}}_{np}, \label{eq:Nyxa}\\
    N_{np}^{\mathrm{YY}, \rm a} &= \mathrm{i}  \,\omega_{s_n}A^{\mathrm{XY},\rm{a}}_{np}, \label{eq:Nyya}
    \end{align}
\noindent where $A^{q,\rm{a}}_{np}$ with $q\in\{\mathrm{XX,XY,YX,YY}\}$ and $\rm a\in\{\text{dip, ex, U}\}$, are the dynamic factors that relate the dynamic magnetization in the $p$-th layer $\tilde{\mathbf{m}}_p[\mathbf{k}]=\tilde{m}_{x_p}\hat x+\tilde{m}_{y_p}\hat y$ to the dynamic magnetic field in the $n$-th layer $\tilde{\mathbf{h}}_{np}^{\text{a}}[\mathbf{k}]$.
More explicitly
\smallskip{}
    \begin{equation}\label{eq:hanpkspace}
    \tilde{\mathbf{h}}_{np}^{\rm a}[\mathbf{k}]=
    \begin{pmatrix}
    \tilde{h}_{x_{np}}^{\rm a}[\mathbf{k}] \\
    \tilde{h}_{y_{np}}^{\rm a}[\mathbf{k}]
    \end{pmatrix}
    =
    \begin{pmatrix}
    A_{np}^{\mathrm{XX},\rm a} & A_{np}^{\mathrm{XY},\rm a} \\
    A_{np}^{\mathrm{YX},\rm a} & A_{np}^{\mathrm{YY},\rm a}
    \end{pmatrix}
    \begin{pmatrix}
    \tilde m_{x_p}[\mathbf{k}] \\
    \tilde m_{y_p}[\mathbf{k}]
    \end{pmatrix},
    \end{equation}

\noindent where the explicit expressions of $A^{q,\text{a}}_{np}$ are given in SM Section III\cite{suplementario}. The dynamic matrix entries of the Zeeman interaction are written similarly, however, since this does not arise from a dynamic magnetic field, we write it appart as

\smallskip{}
 \begin{align}
    N_{np}^{\mathrm{XX}, \rm Ze} &= 0,\label{eq:NxxZe}\\
    N_{np}^{\mathrm{XY}, \rm Ze} &= \mathbf{i} \, \omega_{s_n} \frac{H}{M_{s_n}} \, \, \delta_{p,n}, \label{eq:NxyZe}\\
    N_{np}^{\mathrm{YX}, \rm Ze} &= -\mathbf{i}\, \omega_{s_n} \frac{H}{M_{s_n}} \,  \, \delta_{p,n}, \label{eq:NyxZe}\\
    N_{np}^{\mathrm{YY}, \rm Ze} &= 0. \label{eq:NyyZe}
\end{align}

 The dynamic matrix entries for the interlayer exchange interaction cannot be written in a direct relation to its corresponding dynamic factors $A^{q,\text{int}}_{np}$ (as seen in Eqs. \eqref{eq:Nxxa} - \eqref{eq:Nyya}), and take the form 
\smallskip{}
\begin{align}
& N_{n p}^{\mathrm{XX}, \text { int }}=0,\label{eq:Nnpxxint}  \\
& N_{n p}^{\mathrm{XY}, \text { int }}=i \omega_{s_n}\Bigg(\Bigg(\frac{H_{n+1, n}^{\text {int }}}{M_{s_n}} \delta_{p, n+1}+\frac{H_{n-1, n}^{\text {int }}}{M_{s_n}} \delta_{p, n-1}\Bigg) \nonumber\\
&\,\,\,\,\,\,\,\,\,\,\,\,\,\,\,\,\,\,\,\,\,\,\,\,\,\,\,\,\,\,\,\,\,\,\,\,\,\,\,\,\,\,\,\,\,\,\,\,\,\,\,\,\,\,\,\,\,\,\,\,\,\,\,\,- \Bigg(\frac{H^{\text{int}}_{n,n+1}}{M_{s_n}} + \frac{H^{\text{int}}_{n,n-1}}{M_{s_n}}
\Bigg) \delta_{p,n}\Bigg), \label{eq:Nnpxyint}\\
&N_{n p}^{\mathrm{YX}, \text { int }}=-=i \omega_{s_n}\Bigg(\Bigg(\frac{H_{n+1, n}^{\text {int }}}{M_{s_n}} \delta_{p, n+1}+\frac{H_{n-1, n}^{\text {int }}}{M_{s_n}} \delta_{p, n-1}\Bigg)\nonumber \\
&\,\,\,\,\,\,\,\,\,\,\,\,\,\,\,\,\,\,\,\,\,\,\,\,\,\,\,\,\,\,\,\,\,\,\,\,\,\,\,\,\,\,\,\,\,\,\,\,\,\,\,\,\,\,\,\,\,\,\,\,\,\,\,\,- \Bigg(\frac{H^{\text{int}}_{n,n+1}}{M_{s_n}} + \frac{H^{\text{int}}_{n,n-1}}{M_{s_n}}
\Bigg) \delta_{p,n}\Bigg), \label{eq:Nnpyxint}\\
& N_{n p}^{\mathrm{YY}, \text { int }}=0, \label{eq:Nnpyyint}
\end{align}
\smallskip{}

\noindent  where $ H^{\text{int}}_{np}$ is the strength of the inter-exchange coupling between n'th and p'th layers. This field is given in the SM Section III\cite{suplementario}, however we bring it here as
\smallskip{}
\begin{equation}
    H^{\text{int}}_{np} = - \frac{J_{np}}{ \, \mu_0 \, M_{s_n} \, d}.\label{eq:Hintnpfield}
\end{equation}
\noindent For simplicity, we assume identical sublayer thicknesses $d_n=d$. 

Nonreciprocal dispersion  is quantified by evaluating the eigenfrequencies $\omega_{\pm}=\omega[\pm\mathbf{k}]$ and the resulting frequency shift $\Delta \omega\equiv\omega_+-\omega_-$. A vanishing $\Delta\omega$ corresponds to reciprocal dispersion, while larger values indicate stronger nonreciprocity. Reported frequency shifts in multilayer systems typically reach $5$-$10$~GHz (and up to $12$~GHz) \cite{korber2023}, and are commonly attributed to dipolar interactions. This interpretation is motivated by the fact that only the dipolar matrix elements explicitly depend on the propagation direction, i.e., $N_{np}^{\mathrm{XX},\text{dip}}[\mathbf k]\neq N_{np}^{\mathrm{XX},\text{dip}}[-\mathbf k]$ and $N_{np}^{\mathrm{YY},\text{dip}}[\mathbf k]\neq N_{np}^{\mathrm{YY},\text{dip}}[-\mathbf k]$. However, as we demonstrate below, attributing the frequency shift values solely to dipolar interactions provides an incomplete physical picture. Instead, exchange interaction can play an equally important - or even dominant - role. To elucidate this mechanism, we introduce a frequency-shift dynamic matrix $\mathbbm{W}$ whose eigenvalues correspond to $\Delta\omega$ and whose eigenvectors are $(\tilde{\mathbbm{m}}^+,\tilde{\mathbbm{m}}^-)^T$.

\subsection{Frequency Shift Dynamic Matrix (FSDM)} \label{Frequency Shift Dynamic Matrix}
The frequency-shift dynamic matrix (FSDM) $\mathbbm{W}$ is an operator whose eigenvalues correspond to the frequency shift $\Delta\omega$, while its eigenvectors are the counterpropagating spin-wave oscillation modes $\tilde{\mathbbm{m}}[\pm\mathbf{k}]$. The explicit form of $\mathbbm{W}$ and the derivation of its associated eigenvalue problem are presented in the SM Section I\cite{suplementario}. For completeness, we summarize here the key expressions. Accordingly, the eigenvalue equation defining the FSDM reads
\smallskip{}
\begin{align}
\Delta \omega \,\tilde{\mathbbm{m}}[\mathbf{\pm k}] &= \mathbb{W}[\mathbf{\pm k}] \, \tilde{\mathbbm{m}}[\mathbf{\pm k}], \label{eq:Wposi}
\end{align}

\noindent which constitutes a nonlinear eigenvalue problem, since the operator $\mathbbm{W}$ depends explicitly on the spin-wave eigenmodes $\tilde{\mathbbm{m}}$. Similar nonlinear eigenvalue problems are well known in other branches of physics, such as nonlinear optics and nonlinear quantum mechanics, where they are typically solved using self-consistent methods \cite{ClaesMeerbergen2023,QuantumMechanics_Feng_2011}.
Our goal is not to solve Eq.~\eqref{eq:Wposi} explicitly, but to use the structure of $\mathbbm{W}$, to disentangle and quantify the respective roles of dipolar, interlayer and intralayer exchange (again, only the decomposition of the isotropic exchange), and uniaxial anisotropy interactions in the frequency shift of nonreciprocal spin-wave dispersion relations.
Accordingly, the FSDM can be expressed as
\begin{equation}
\mathbb{W}=\mathbb{W}^{\text{dip}}+\mathbb{W}^{\text{int}}+\mathbb{W}^{\text{ex}}+\mathbb{W}^{\text{U}}+\mathbb{W}^{\text{Ze}}
\label{eq:FSDMContributions}
\end{equation}

To easier the mathematical results, we use the notation $\mathbb{X}^{\sigma,\pm}\equiv\mathbb{X}^{\sigma}[\mathbf{\pm k}]$, thus, each interaction contribution to the FSDM can be written explicitly as

\smallskip{}
\begin{align}
\mathbb{W}^{\sigma,+} &= \mathbb{N}^{\sigma}[\mathbf{+ k}] - \tilde{\mathbb{N}}^{\sigma}[\mathbf{-k}], \label{eq:wpositivo} \\
\mathbb{W}^{\sigma,-}&= \tilde{\mathbb{N}}^{\sigma}[\mathbf{+k}] - \mathbb{N}^{\sigma}[\mathbf{-k}], \label{eq:wnegativo}
\end{align}

\noindent where $\tilde{\mathbb{N}}^{\sigma,\pm}$ is the transformed dynamic matrix of the $\sigma$th interaction defined by
\smallskip{}
\begin{align}
    & \tilde{\mathbb{N}}^{\sigma,+}\equiv  \tilde{\mathbb{O}}\, \left(\mathbb{N}^{\sigma,-}\right)\, \tilde{\mathbb{O}}^{-1},\\
    &\tilde{\mathbb{N}}{\sigma,-} \equiv  \tilde{\mathbb{O}}^{-1}\, \left(\mathbb{N}^{\sigma,+}\right)\, \tilde{\mathbb{O}},
\end{align}

\noindent and $\tilde{\mathbb{O}}$ is an operator that transforms the spin-wave mode $\tilde{\mathbbm{m}}[\mathbf{- k}]$ into $\tilde{\mathbbm{m}}[\mathbf{+ k}]$ (i.e., $\tilde{\mathbbm{m}}[\mathbf{+ k}]=\tilde{\mathbb{O}}\tilde{\mathbbm{m}}[\mathbf{- k}]$). This operator is defined as:
\smallskip{}
\begin{equation}
\tilde{\mathbb{O}} \;=\;
\begin{pmatrix}
\tilde{\mathbbm{a}} \, \mathbb{R}_{11} & \tilde{\mathbbm{a}} \, \mathbb{R}_{21} \\[6pt]
\tilde{\mathbbm{a}} \, \mathbb{R}_{12} & \tilde{\mathbbm{a}} \, \mathbb{R}_{22}
\end{pmatrix}, \label{eq:matrixO}
\end{equation}

\noindent where  $\tilde{\mathbbm{a}}$ and $\mathbbm{R}_{ij}$ with $i,j = 1,2$, are $N \times N$ diagonal matrices. The n'th diagonal entry of $\tilde{\mathbbm{a}}$ is   $\alpha_n=\left\lVert \tilde{\mathbf{m}}_{n}^{+} \right\rVert/\left\lVert \tilde{\mathbf{m}}_{n}^{-} \right\rVert$ with $\tilde{\mathbf{m}}_{n}^{\pm}=\tilde{\mathbf{m}}_{n}[\pm\mathbf{k}]$, and the n'th diagonal entry of $\mathbbm{R}_{ij}$, denoted by $R^n_{ij}$, is defined by the entries of the 2x2 rotational matrix
\smallskip{}
\begin{equation}
\mathbb{R}_n=
\begin{bmatrix}
R_{11}^{n} &
R_{12}^{n} \\[4pt]
R_{21}^{n} &
R_{22}^{n}
\end{bmatrix},\label{eq:Riequation}
\end{equation}

\noindent that rotates the unit vectors $\mathbf{\hat{u}}_n^{\pm}=\tilde{\mathbf{m}}_n^{\pm}/\left\lVert \tilde{\mathbf{m}}_{n}^{\pm} \right\rVert$ as $\mathbf{\hat{u}}_n^+=\mathbb{R}_n \mathbf{\hat{u}}_n^-$. The entries $R^n_{ij}$ are in general complex, i.e., $R_{ij}^n=\text{Re}[R_{ij}^n]+\mathrm{i}\, \text{Re}[R_{ij}^n]$, with $\mathrm{i}$ the imaginary complex unit. 

Because $\mathbb{R}_n$ is a unitary operator, it is sufficient to analyze the frequency shift with $\mathbbm{W}^{\sigma,+}$ (the analysis with $\mathbbm{W}^{\sigma,-}$ is equivalent). Therefore, we write $\mathbbm{W}^{\sigma,+}$ as
\smallskip{}
\begin{equation}
\mathbbm{W}^{\sigma,+} =
\begin{pmatrix}
\mathbbm{W}_{\mathrm{XX}}^{\sigma,+} & \mathbbm{W}_{\mathrm{XY}}^{\sigma,+} \\
\mathbbm{W}_{\mathrm{YX}}^{\sigma,+} & \mathbbm{W}_{\mathrm{YY}}^{\sigma,+}
\end{pmatrix},
\label{eq:FSDMBlocks}
\end{equation}

\noindent focusing on the DE configuration, where  

\begin{equation}
\mathbbm{W}_{q}^{\sigma,+}=\mathbb{N}^{\sigma,+}_{q} -\tilde{\mathbb{N}}^{\sigma,-}_{q} 
\label{eq:FSDMBLOCKS}
\end{equation}

\noindent with $q\in\{\mathrm{XX, XY, YX, YY}\}$. In this case one can show that $\mathbbm{R}_{11}$ and $\mathbbm{R}_{12}$ are either pure real or imaginary matrices, i.e., $(R_{11}^n)^*=\pm R_{11}^n$ and $(R_{12}^n)^*=\pm R_{12}^n$, but not simultaneously, leading to two mutually exclusive cases. The resulting expressions for the blocks $\tilde{\mathbbm{N}}_{q}^{\sigma+}$, are
\smallskip{}

\begin{itemize}
\item For dipolar interaction:
{\small
\begin{align}
\tilde{\mathbbm{N}}_{\mathrm{XX}}^{\mathrm{dip},+}&=\mathbbm{A}_{\mathrm{XX}}^{\mathrm{dip}}+\mathbbm{A}_{\mathrm{XY}}^{\mathrm{dip}}, \label{eq:TNXXdip}\\
\tilde{\mathbbm{N}}_{\mathrm{YY}}^{\mathrm{dip},+}&=-\mathbbm{A}_{\mathrm{XX}}^{\mathrm{dip}}+\mathbbm{A}_{\mathrm{XY}}^{\mathrm{dip}}, \label{eq:TNYYa}\\
\tilde{\mathbbm{N}}_{\mathrm{XY}}^{\mathrm{dip},+}&=\mathbbm{P}_{\mathrm{XX}}^{\mathrm{dip}}+\mathbbm{P}_{\mathrm{XY}}^{\mathrm{dip}}+\mathbf{i}\,\mathbbm{w}\left(\mathbbm{R}_{12}\right)^{~2}, \label{eq:TNXYdip}\\
\tilde{\mathbbm{N}}_{\mathrm{YX}}^{\mathrm{dip},+}&=-\mathbbm{P}_{\mathrm{XX}}^{\mathrm{dip}}+\mathbbm{P}_{\mathrm{XY}}^{\mathrm{dip}}+\mathbf{i}\,\mathbbm{w}\left(\mathbbm{R}_{11}^{\dagger}\right)^{~2},\label{eq:TNYXdip}
\end{align}
}
\noindent where $\mathbbm{A}_{q}^{\mathrm{dip}}$ and $\mathbbm{P}_{q}^{\mathrm{dip}}$ are defined as

{\small
\begin{align}
\mathbbm{A}_{\mathrm{XX}}^{\mathrm{dip}}&\equiv-\mathbb{R}_{11}\left\langle\mathbbm{N}_{\mathrm{XX}}^{\mathrm{dip},+}\right\rangle \mathbbm{R}_{11}^{\dagger}+\mathbbm{R}_{12}\left\langle\mathbbm{N}_{\mathrm{XX}}^{\mathrm{dip},+}\right\rangle \mathbbm{R}_{12}^{\dagger},\\
\mathbbm{A}_{\mathrm{XY}}^{\mathrm{dip}}&\equiv\mathbbm{R}_{12}\left\langle\mathbbm{N}_{\mathrm{XY}}^{\mathrm{dip},+}\right\rangle \mathbbm{R}_{11}^{\dagger}+\mathbbm{R}_{11}\left\langle\mathbbm{N}_{\mathrm{XY}}^{\mathrm{dip},+}\right\rangle \mathbbm{R}_{12}^{\dagger}-\mathrm{i}\, \mathbbm{w}\mathbbm{R}_{12} \mathbbm{R}_{11}^{\dagger},\\
\mathbbm{P}_{\mathrm{XX}}^{\mathrm{dip}}&\equiv\mathbbm{R}_{12}\left\langle\mathbbm{N}_{\mathrm{XX}}^{\mathrm{dip},+}\right\rangle \mathbbm{R}_{11}+\mathbbm{R}_{11}\left\langle\mathbbm{N}_{\mathrm{XX}}^{\mathrm{dip},+}\right\rangle \mathbbm{R}_{12},\\
\mathbbm{P}_{\mathrm{XY}}^{\mathrm{dip}}&\equiv\mathbbm{R}_{11}\left\langle\mathbbm{N}_{\mathrm{XY}}^{\mathrm{dip},+}\right\rangle \mathbbm{R}_{11}-\mathbbm{R}_{12}\left\langle\mathbbm{N}_{\mathrm{XY}}^{\mathrm{dip},+}\right\rangle \mathbbm{R}_{12}+\mathrm{i}\, \mathbbm{w}\mathbbm{R}_{12} \mathbbm{R}_{12}
\end{align}
}

\item For interlayer exchange, intralayer exchange, uniaxial anisotropy, and Zeeman interactions:

{\small
\begin{align}
\tilde{\mathbb{N}}_{\mathrm{XX}}^{\mathrm{a},+}&=\mathbbm{A}_{\mathrm{XY}}^{\mathrm{a}}\label{eq:TNXXa}\\
\tilde{\mathbb{N}}_{\mathrm{XY}}^{\mathrm{a},+}&=\mathbbm{P}_{\mathrm{XY}}^{\mathrm{a}}\label{eq:TNXYa}\\
\tilde{\mathbb{N}}_{\mathrm{YY}}^{\mathrm{a},+}&=-\mathbbm{A}_{\mathrm{XY}}^{\mathrm{a}}\label{eq:TNYYa}\\
\tilde{\mathbb{N}}_{\mathrm{YX}}^{\mathrm{a},+}&=-\mathbbm{P}_{\mathrm{XY}}^{\mathrm{a}}\label{eq:TNYXa}
\end{align}
}

\noindent where $\mathbbm{A}_{\mathrm{XY}}^{\mathrm{a}}$ and $\mathbbm{P}_{\mathrm{XY}}^{\mathrm{a}}$ for $\mathrm{a}=\mathrm{int}$  are defined as

{\small
\begin{align}
\mathbbm{A}_{\mathrm{XY}}^{\mathrm{int}}&\equiv-\mathbb{R}_{12}\left\langle \mathbbm{N}_{\mathrm{XY}}^{\text{int},+}\right\rangle \mathbb{R}_{11}^{\dagger }+\mathbb{R}_{11}\left\langle \mathbbm{N}_{\mathrm{XY}}^{\text{int},+}\right\rangle\mathbb{R}_{12}^{\dagger},\label{eq:AXXint}\\
\mathbbm{P}_{\mathrm{XY}}^{\mathrm{int}}&\equiv\mathbb{R}_{11}\left\langle \mathbbm{N}_{\mathrm{XY}}^{\text{int},+}\right\rangle\mathbb{R}_{11}+\mathbb{R}_{12}\left\langle \mathbbm{N}_{\mathrm{XY}}^{\text{int},+}\right\rangle\mathbb{R}_{12},\label{eq:PYXint}
\end{align}
}

\noindent and $\mathbbm{A}_{\mathrm{XY}}^{\mathrm{a}}$ and $\mathbbm{P}_{\mathrm{XY}}^{\mathrm{a}}$ for $\mathrm{a}\in\{\mathrm{ex, U, Ze}\}$  are defined as

{\small
\begin{align}
\mathbbm{A}_{\mathrm{XY}}^{\mathrm{a}}&\equiv-\left(\mathbb{R}_{11} \mathbb{R}_{12}^{\dagger }-\mathbb{R}_{12} \mathbb{R}_{11}^{\dagger}\right)\mathbbm{N}_{\mathrm{XY}}^{\mathrm{a},+},\label{eq:AXXa}\\
\mathbbm{P}_{\mathrm{XY}}^{\mathrm{a}}&\equiv\left(\mathbb{R}_{11} \mathbb{R}_{11}+\mathbb{R}_{12} \mathbb{R}_{12}\right)\mathbbm{N}_{\mathrm{XY}}^{\mathrm{a},+},\label{eq:AXYa}
\end{align}
}
\end{itemize}

\noindent Here, $\langle \mathbbm{N}_{q}^{\sigma,+}\rangle\equiv \tilde{\mathbbm{a}}^{-1}\mathbbm{N}_{q}^{\sigma,+}\tilde{\mathbbm{a}}$ is known as a simmilarity transformation of $\mathbbm{N}_{q}^{\sigma,+}$ by the diagonal matrix $\tilde{\mathbbm{a}}$, and $\mathbbm{w}$ is a $N\times N$ diagonal matrix with the n'th diagonal entry given as $[\mathbbm{w}_n]=\omega_{s_n}\delta_{np}$.

The entries $R_{11}^n$ and $R_{12}^n$ can be determined in basis of the dynamic magnetization equation at the n'th layer $\mathbf{m}_{n}[t; \mathbf{k}, \omega]=m_{x_n}[t; \mathbf{k}, \omega] \hat x+m_{y_n}[t; \mathbf{k}, \omega] \hat y$ when evaluated at $x=0$ and $z=0$ (see SM Section II\cite{suplementario}). Its components are given as 

\smallskip{}
\begin{align}
m_{i_n}[t;\mathbf{k},\omega] &= \text{Re}[\tilde{m}_{i_n}[\mathbf{k}]] \cos[\omega t] - \text{Im}[\tilde{m}_{i_n}[\mathbf{k}]] \sin[\omega t].\label{eq:mynt}
\end{align}

\noindent  with $i\in\{x,y\}$ and $\tilde{\mathbf{m}}_n=\tilde{m}_{x_n}\hat x+\tilde{m}_{y_n}$. Since the dynamic magnetization can be written in terms of the geometric parameters of spin wave orbit  such as the eccentricity $\epsilon_n$, tilting angle $\varphi_{n}^{0}$, and phase $\tau_n$, Fig. \ref{fig:system}c), thus one can write $R_{11}^n$ and $R_{12}^n$ in terms of these parameters which can be expressed as  

\smallskip{}
{\small
\begin{align}
    \text{Re}[{R}_{11}^{n}]&=r_0^n 
\Big( \Big(\sqrt{(1-\epsilon_{n}^{\!-~2})(1-\epsilon_{n}^{\!+~2})} + 1\Big) 
 \cos[\Delta\varphi_{n}^{0}]\cos[\Delta\tau_{n}] \nonumber\\
&\quad - \Big(\sqrt{1-\epsilon_{n}^{\!+~2}} + \sqrt{1-\epsilon_{n}^{\!-~2}}\Big) 
\sin[\Delta\varphi_{n}^{0}]\sin[\Delta\tau_{n}]\Big),\label{eq:ReR11} \\
\text{Re}[{R}_{12}^{n}]&=r_0^n 
\Big(\Big( \sqrt{1 - \epsilon_{n}^{\!+~2}} + \sqrt{1 - \epsilon_{n}^{\!-~2}} \Big)
\cos[\Delta\varphi_{n}^{0}]
\sin[\Delta\tau_{n}] \nonumber \\
&+ \Big( \sqrt{(1 - \epsilon_{n}^{\!+~2})(1 - \epsilon_{n}^{\!-~2})} + 1 \Big) \sin[\Delta\varphi_{n}^{0}] \cos[\Delta\tau_{n}]\Big),\\
\text{Im}[R_{11}^n]&=r_0^n\Big( \Big(1 - \sqrt{(1-\epsilon_{n}^{\!-~2})(1-\epsilon_{n}^{\!+~2})}\Big) 
\cos[\bar{\varphi}_{n}^{0}] \sin[\Delta\tau_{n}]\nonumber \\
&\quad + \Big(\sqrt{1-\epsilon_{n}^{\!+~2}} - \sqrt{1-\epsilon_{n}^{\!-~2}}\Big) 
\sin[\bar{\varphi}_{n}^{0}]\cos[\Delta\tau_{n}]  \Big),\\
\text{Im}[R_{12}^n]&=r_0^n\Big(
\Big( -\sqrt{1 - \epsilon_{n}^{\!+~2}} + \sqrt{1 - \epsilon_{n}^{\!-~2}} \Big)
\cos[\bar{\varphi}_{n}^{0}]\cos[\Delta\tau_{n}] \nonumber \\
&\qquad +
\Big( \sqrt{(1 - \epsilon_{n}^{\!+~2})(1 - \epsilon_{n}^{\!-~2})} - 1 \Big)\sin[\bar{\varphi}_{n}^{0}]\sin[\Delta\tau_{n}]
\Big)\label{eq:ImR12}.
\end{align}
}

\noindent where $r_0^n\equiv1/\sqrt{(2-\epsilon_{n}^{\!-~2})(2-\epsilon_{n}^{\!+~2})}$, $\Delta \tau_n=\tau_n^--\tau_n^+$ is a phase shift, $\Delta \varphi_n^0=\varphi_n^{0-}-\varphi_n^{0+}$ is a tilting angle shift, and $\bar\varphi_n^0=\varphi_n^{0+}+\varphi_n^{0-}$. Note the notation $x^{\pm}=x[\pm \mathbf{k}]$ apply to all SWs orbit parameters, i.e., with $x$ running on $\epsilon_n$, $\varphi_n^0$ and $\tau_n$, which are given as
\smallskip{}
\begin{align}
    &\epsilon_n=\sqrt{1-\left(\frac{b_n}{a_n}\right)^2},\label{eq:eccentricity}\\
    &\tan\Big[2 \, \varphi_{n}^{0}\Big] = 
- \frac{\operatorname{Im}\big[\tilde{m}_{x_n}^2\big] + \operatorname{Im}\big[\tilde{m}_{y_n}^2\big]}
       {\operatorname{Re}\big[\tilde{m}_{x_n}^2\big]+ \operatorname{Re}\big]\tilde{m}_{y_n}^2\big]},\label{eq:tilting}\\
    &\tau_n = \arctan\Bigg[\frac{\operatorname{Re}[\tilde{m}_{y_n}]}{\operatorname{Re}[\tilde{m}_{x_n}]}\Bigg] - \varphi_n^0.\label{eq:phaseshift}
\end{align}

These relations show that the FSDM entries depend on the orbit geometry at each sublayer, and hence on how the mode profiles at $\pm\mathbf{k}$ differ along the thickness. As we will see, this dependence is crucial for understanding when the dipolar interaction alone accounts for the nonreciprocal frequency shift and when the exchange interactions, in particular interlayer exchange, become dominant.

\section{Results and Discussion}
In this section, we use the frequency-shift dynamic matrix (FSDM) to identify which magnetic interactions dominate the nonreciprocal frequency shift in multilayer systems. We first analyze the role of each interaction in the FSDM by comparing the entries among the dynamic matrix blocks of the interactions. Then, we analyze the structure of the FSDM for the particular case where the dipolar interaction dominates the frequency shift of counterpropagating modes, and for the general case in which the interlayer exchange interaction dominates, even in systems where the frequency shift is usually attributed solely to dipolar fields. Finally, we relate these findings to the geometry of the spin-wave orbits and the rotation matrices that enter the FSDM.

\subsection{Interaction strengths and their role to the frequency shift}

To identify the dominant interactions in the SWs frequency shift, let's first analyze how each interaction is weighted in the FSDM, $\mathbbm{W}^+$. From Eqs. \eqref{eq:wpositivo} and \eqref{eq:TNXXdip}-\eqref{eq:AXYa}, one can see that $\mathbbm{R}_{11}$, $\mathbbm{R}_{12}$ and their adjoints play as weights of the dynamic matrix blocks $\mathbbm{N}_{\mathrm{XX}}^{\mathrm{dip}}$  and $\mathbbm{N}_{\mathrm{XY}}^{\sigma}$ in the FSDM blocks $\mathbbm{W}_{q}^{\sigma,+}$. Since the norm of the entries of $\mathbbm{R}_{11}$ and $\mathbbm{R}_{12}$ are delimited between 0 and 1, thus the dynamic matrices are equally weighted by them in the FSDM. Accordingly, the contribution degree of each interaction in the FSDM can be evaluated by comparing the entries among the dynamic matrix blocks of the interactions. So, from Eqs. \eqref{eq:Nxxa}-\eqref{eq:Nyya} and Eqs. \eqref{eq:Nnpxxint}-\eqref{eq:Nnpyyint}, we introduce the so-called ``\textit{the strength}'' of each interaction as follows
\smallskip{}
\begin{align}
    &S_{\text{dip}}^{(1)}[k]=-\frac{1}{k d/2}\sinh ^2\left[k\frac{d}{2}\right]e^{-k d}, \label{equation:f1dip}\\
    &S_{\text{dip}}^{(2)}[k]=-\frac{1}{k d/2}\sinh\left[k\frac{d}{2}\right]e^{-k d/2},\label{equation:f2dip}\\
    &S_{\text{ex}_n}=l_{\text{ex}_n}|\mathbf{k}|^2,\label{equation:fex}\\
    &S_{\text{int}_n}=\frac{H_{n+1,n}^{\text{int}}}{M_{s_n}},\label{equation:fint}\\
    &S_{\text{Ze}_n}=\frac{H_0}{M_{s_n}},\label{equation:fZe}\\
    &S_{\text{U}_n}=\frac{H_n^{\text{U}}}{M_{s_n}}, \label{equation:fU}
\end{align}

\noindent where $S_{\text{dip}}^{(1)}[k]$ and $S_{\text{dip}}^{(2)}[k]$ define the strength of the  dynamic dipolar blocks $\mathbbm{N}_{\mathrm{XX}}^{\text{dip}}$ and $\mathbbm{N}_{\mathrm{XY}}^{\text{dip}}$, and  $S_{\mathrm{a}_n}$ the strength at the n'th sublayer of the  dynamic $\mathrm{a}$'th block $\mathbbm{N}_{\mathrm{XY}}^{\mathrm{a}}$ with $\mathrm{a}\in\{\text{int, ex, U, Ze}\}$.  

To compare the field strengths defined above, we evaluate them for a generic multilayer with magnetic parameters spanning a wide range across the sublayers. Hence, by adopting the notation of the $X$ magnetic parameter ranging as $X\in[X_{\text{min}},X_{\text{max}}]$, we have: $A_n\in[10^{-12},10^{-10}]\,\mathrm{J/m}$,
$M_{s_n}\in[700,1800]\times 10^{3}\,\mathrm{A/m}$,
$K_n^{\mathrm{U}}\in[0,10^{6}]\,\mathrm{J/m^3}$,
$d=1\,\mathrm{nm}$, and interlayer exchange $J_{n,p}=2A_n/d$. From these we obtain
$l_{\mathrm{ex}_n}\in[0.7,18]\,\mathrm{nm}$,
$H_{n,p}^{\mathrm{int}}\in[0.468,113.69]\times 10^{6}\,\mathrm{A/m}$,
$H_n^{\mathrm{U}}\in[0,0.274]\times 10^{6}\,\mathrm{A/m}$ and
$H\in[0,8]\times 10^{6}\,\mathrm{A/m}$. The corresponding strength ranges are then
$S_{\mathrm{int}_n}\in[0.28,162.4]$,
$S_{\mathrm{U}_n}\in[0,0.323]$, and
$S_{\mathrm{Ze}_n}\in[0,1]$, while $S_{\mathrm{ex}_n}\propto|\mathbf{k}|^{2}$ through $l_{\mathrm{ex}_n}$.

\begin{figure}[th!]
    \centering
    \includegraphics[scale=0.29]{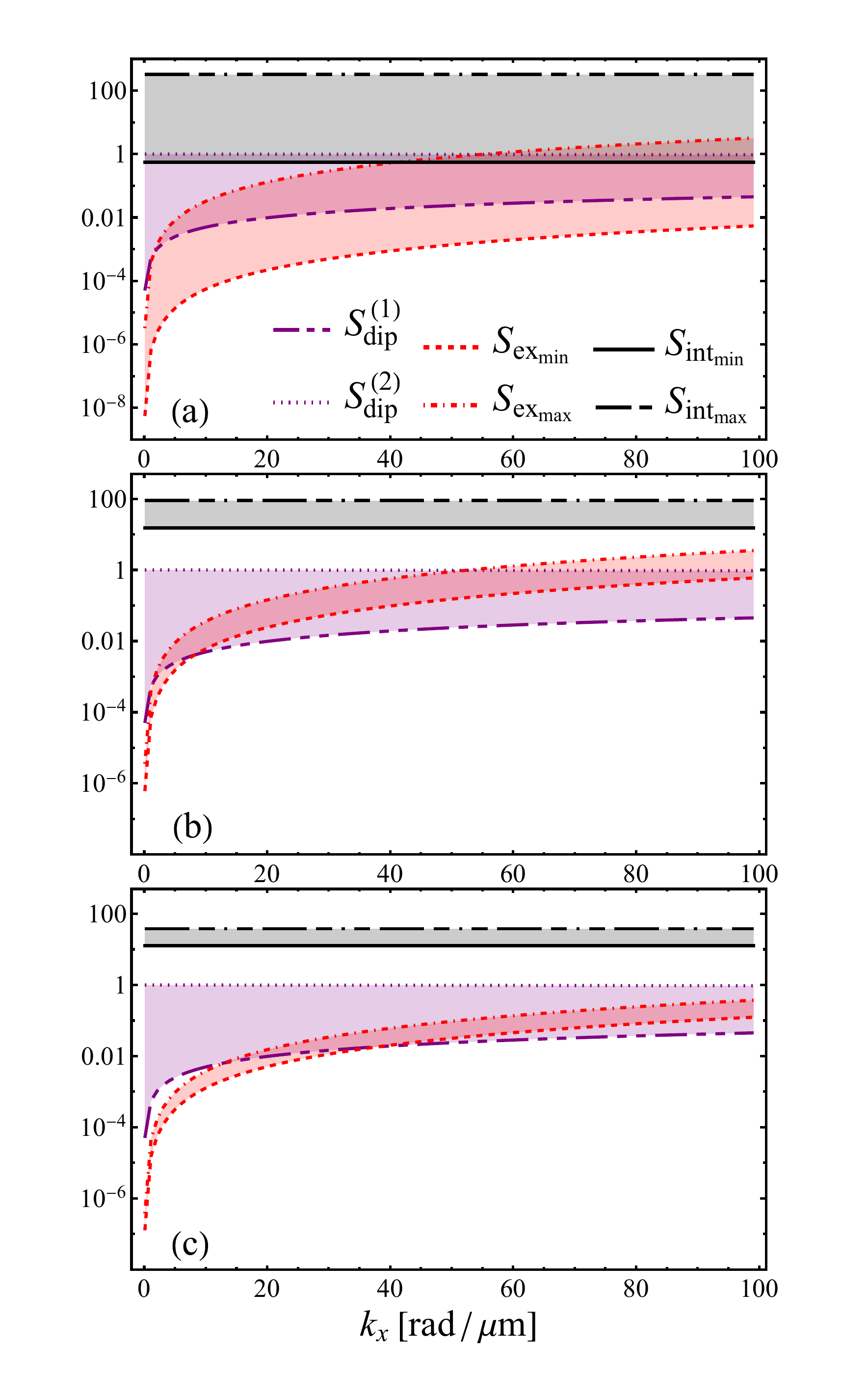} 
    \caption{Dipolar interaction strength ($S_{\text{dip}}^{(1)}$ and $S_{\text{dip}}^{(2)}$), and maximum and minimum strengths of the interlayer exchange interaction ($S_{\text{int}_{\text{min}}},S_{\text{int}_{\text{max}}}$) and intralayer exchange interaction ($S_{\text{ex}_{\text{min}}}, S_{\text{ex}_{\text{max}}}$), as a function of the wave vector $k_{x}$. (a) General multilayered film, (b) graded magnetization NiFe layer, and (c) Magnonic diode CoFeB/NiFe bilayer.}
    \label{fig:InteractionStrengths}
\end{figure}

Fig. \ref{fig:InteractionStrengths}(a) shows these interaction strengths as a function of the wave vector $k=k_x$ over the full parameter ranges, except for the uniaxial anisotropy and Zeeman interactions, whose maximum and minimum values remain constant. Despite the wide range of parameters considered, the interlayer exchange interaction strength (gray region) is generally larger than the other interaction strengths. (i) The dipolar interaction strength (purple region) competes most strongly with the interlayer exchange interaction, sharing only a limited overlapping range between approximately 0.8 and 1 across all wave vector values. (ii) The intralayer exchange interaction strength (red region) competes only weakly with the interlayer exchange strength, and only for wave vectors $k_x>30\text{ rads/$\mu$m}$, where a small overlap between the gray and red regions appears. (iii) The Zeeman interaction and uniaxial anisotropy strengths compete negligibly with the interlayer exchange interaction, as their regions show almost no overlap with it. This is because the maximum strengths of these interactions ($S_{\text{Ze}_{\text{max}}}=1$ and $S_{\text{U}_{\text{max}}}=0.323$)  are of the same order as the minimum interlayer exchange strength ($S_{\text{int}_{\text{min}}}=0.28$).

To connect this generic analysis to realistic systems, we evaluate the strengths for two experimentally relevant cases from Ref.~\citep{GrassiDIODO2020,Gallardo2019}. The first system $\textit{``The magnonic diode"}$ (MD) \cite{GrassiDIODO2020} considers a CoFeB/NiFe bilayer, where the CoFeB (NiFe) layer have a homogeneous magnetization, stiffness constant, thickness and uniaxial anisotropy constant as $M_s^{\text{CoFeB}}$ = 1270 $\text{kA/m}$ ($M_s^{\text{NiFe}}$ = 845 $\text{kA/m}$), $A_{\text{CoFeB}}$ = 17 $\text{pJ/m}$ ($A_{\text{NiFe}}$ = 12.8 $\text{pJ/m}$), $d_{t,\text{CoFeB}}=25$ nm ($d_{t,\text{NiFe}}=25$ nm) and $K_{\text{CoFeB}}=0$ ($K_{\text{NiFe}}=0$), respectively, with an applied magnetic field $\mu_0 H=30$ mT. The second system $\textit{``The graded magnetization layer"}$ (GML) \cite{Gallardo2019} considers a planar NiFe layer with graded saturation magnetization along the thickness whose profile changes from $M_s$ = 800 $\text{kA/m}$ to $M_s$ = 1600 $\text{kA/m}$, a thickness $d_{t,\text{NiFe}}=60$ nm and a stiffness constant $A_{\text{NiFe}}$ = 11 $\text{pJ/m}$. For this system, the applied field is $\mu_0 H=1.5 $ mT along the saturation magnetization direction. Their corresponding interaction strength results in Fig \ref{fig:InteractionStrengths}(b) and (c), show that in both structures the interlayer exchange strength remains the largest contribution across the full $k_x$ range. This is noteworthy because, on the one hand, in the original works the frequency shift was attributed primarily to dipolar effects, whereas the strength analysis and the the FSDM analysis shown next clearly indicate that interlayer exchange must play a major role, and on the other hand, they have been the basis for numerous forthcoming literature that follows a similar dipolar based explanation trend.

\subsection{Frequency shift dynamic matrix analysis}

Our results from previous section  reveal the dominant contribution of the interlayer exchange strength to the entries of the FSDM, indicating that this interaction warrants a more detailed examination of its role in shaping the frequency shift of nonreciprocal dispersion relations. Motivated by this observation, we proceed with an analysis that enables a separate and quantitative evaluation of the contributions of each magnetic interaction to the frequency shift in a generalized case and in real systems as the magnonic diode (MD) and graded-magnetization layer (GML) systems.

\begin{figure}
\includegraphics[width=8.2cm]{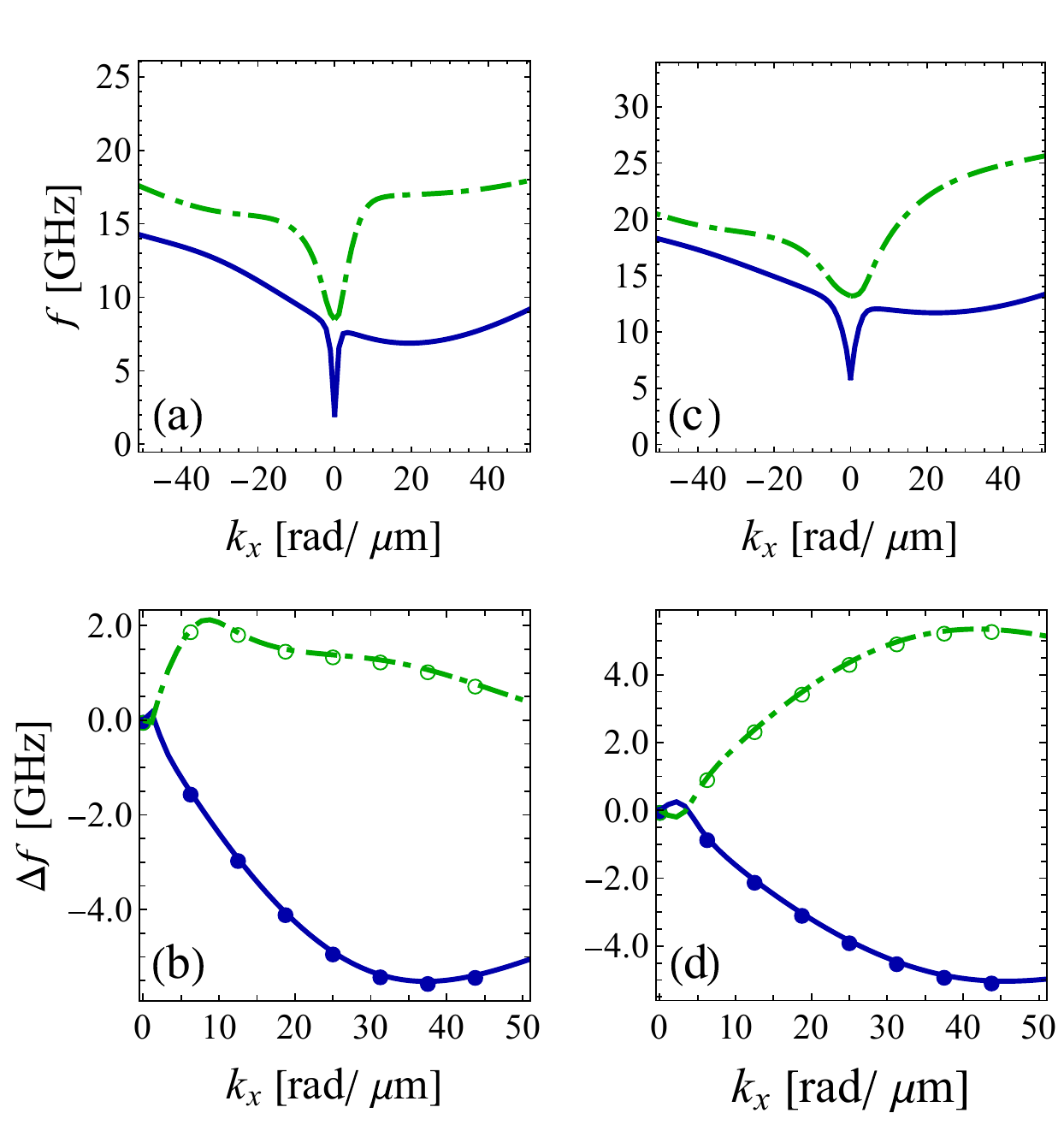}
\caption{(a) Dispersion relation $f$ v.s $k$ and (b) frequency shift $\Delta f\equiv f[k]-f[-k]$ of a graded magnetization NiFe multilayer. (c) and (d) present the same quantities of a CoFeB/NiFe bilayer. Blue-circle and green open circle points in (b) and (d) correspond to the frequency shift calculated with the Frequency Shift Dynamic Matrix, whereas the full blue line and dot-dashed green line correspond to the frequency shift calculated with the Dynamic Matrix. The blue and dot-dashed green curves show the frequency and frequency shift of the fundamental and first-order spin wave modes. The planar NiFe layer has a graded magnetization saturation along the thickness whose profile changes from $M_s$ = 800 $\text{kA/m}$ to $M_s$ = 1600 $\text{kA/m}$, a thickness $d_{t,\text{NiFe}}=60$ nm and a stiffness constant $A_{\text{NiFe}}$ = 11 $\text{pJ/m}$. For this system, the applied field is $\mu_0 H=1.5 $ mT along the saturation magnetization direction. The CoFeB/NiFe bilayer has a saturation magnetization, stiffness constant, thickness and uniaxial anisotropy constant as $M_s^{\text{CoFeB}}$ = 1270 $\text{kA/m}$ ($M_s^{\text{NiFe}}$ = 845 $\text{kA/m}$), $A_{\text{CoFeB}}$ = 17 $\text{pJ/m}$ ($A_{\text{NiFe}}$ = 12.8 $\text{pJ/m}$), $d_{t,\text{CoFeB}}=25$ nm ($d_{t,\text{NiFe}}=25$ nm) and $K_{\text{CoFeB}}=0$ ($K_{\text{NiFe}}=0$), respectively, with an applied magnetic field $\mu_0H=30$ mT.}
\label{fig:disp_energi_CoFeB_NiFe_inte}
\end{figure}

Accordingly, we first reproduce the dispersion relation and frequency shift using the dynamic matrix and the FSDM of the MD and the GML. Fig. \ref{fig:disp_energi_CoFeB_NiFe_inte}(a) and (d), reproduce the dispersion relation presented in the MD and GML references, respectively, using the eigensystem equation Eq. \eqref{eq:LINEALIZA_LL}, with dynamic matrix entries shown in Eqs. \eqref{eq:Nxxa}-\eqref{eq:Nyya} and Eqs. \eqref{eq:Nnpxxint}-\eqref{eq:Nnpyyint}. The frequency shift $\Delta f=1/(2\pi)\Delta \omega$ is calculated by the two different methods described here, continuous (fundamental SW mode) and dashed (first SW mode) lines with the dynamic matrix approach, and full and empty dots with the FSDM eigensystem Eq. \ref{eq:Wposi}, as seen in Figs. \ref{fig:disp_energi_CoFeB_NiFe_inte}(b) and (e). This shows the expected consistency of the FSDM approach for calculating the frequency shift. 

In the following, we demonstrate the role of each contribution on the frequency shift from the FSDM in two separate cases. The first one is the particular case where the dipolar interaction is solely responsible for the non-reciprocity, and the general case where the interlayer exchange interaction is the dominating one, where the other interactions contribute but as perturbative terms in the frequency shift. 

\subsubsection{Particular case: Dipolar dominated frequency shift}

Dipolar interaction is the unique one responsible for the non-reciprocal dispersion relation only in such cases of identical counterpropagating modes at a given wave vector magnitude, i.e.,  

\begin{equation}
\tilde{\mathbf{m}}_n^{+}=\tilde{\mathbf{m}}_n^{-}.
\end{equation}

In terms of the orbit parameters, it means that the eccentricities, tilting angles, and phases coincide for $\pm\mathbf{k}$ in every sublayer ($\epsilon_n^{+}=\epsilon_n^{-}$, $\Delta\varphi_n^0=0$, $\Delta\tau_n=0$). Under these conditions the local rotation matrices reduce to the identity, $\mathbb{R}_n=\mathbb{I}^{2\times 2}$ for all $n$, such that $\mathbbm{R}_{11}=\mathbbm{R}_{22}=\mathbbm{I}^{N\times N}$ and $\mathbbm{R}_{12}=\mathbbm{R}_{21}=0$, and the similarity operator reduces to $\tilde{\mathbbm{a}}=\mathbb{I}$. Consequently, the FSDM in Eq.~\eqref{eq:wpositivo} simplifies to
\smallskip{}

\begin{equation}
\mathbbm{W}^{+}
=\mathbbm{W}^{\mathrm{int},+}=\mathbbm{N}^{\mathrm{dip}}[+\mathbf{k}]-\mathbbm{N}^{\mathrm{dip}}[-\mathbf{k}],
\label{eq:FSDMDipDominated}
\end{equation}

\noindent demonstrating that the frequency shift $\Delta \omega$ is determined solely by the dipolar interaction. This case coincides with the common assumption in the literature that nonreciprocity is purely dipolar in origin. However, Eqs.~\eqref{eq:FSDMDipDominated} also make clear that this conclusion is valid only when the mode profiles are identical for $\pm\mathbf{k}$. As soon as both counterpropagating modes are different, the interlayer exchange interaction dominates, as seen next. 

\subsubsection{General case: Interlayer exchange dominated frequency shift}

Let's evaluate the case of two different counterpropagating modes at a given wave vector in each sublayer. Two cases can be analyzed: the first, consisting of identical normalized profiles, and the general case, in which both modes differ in norm and magnitude. The first case can be described as 

\begin{equation}
\frac{\tilde{\mathbf{m}}_n^{+}}{\lVert \tilde{\mathbf{m}}_n^{+}\rVert} = \frac{\tilde{\mathbf{m}}_n^{-}}{\lVert \tilde{\mathbf{m}}_n^{-}\rVert},
\qquad n=1,\dots,N.
\end{equation}

\noindent wherein eccentricities, tilting angles, phases, and matrices $\mathbbm{R}_{11}$ and $\mathbbm{R}_{12}$ take the same values as for the particular case of identical counterpropagating modes. The only difference is that the similarity matrix $\tilde{\mathbbm{a}}$ is not the identity matrix. Consequently, the FSDM in Eq.~\eqref{eq:FSDMContributions}, Eq.~\eqref{eq:wpositivo} and Eqs.~\eqref{eq:TNXXdip}-\eqref{eq:TNYXa}  simplifies to 
\smallskip{}

\begin{equation}
\mathbbm{W}^{+}=\mathbbm{W}^{\mathrm{dip},+}+\mathbbm{W}^{\mathrm{int},+},
\label{eq:FSDMDipInterDominated}
\end{equation}

\noindent so that the frequency shift $\Delta\omega$ is determined by the dipolar and interlayer interactions. All other interactions appear identically in $\mathbbm{N}[\mathbf{k}]$ and $\mathbbm{N}[-\mathbf{k}]$ and cancel out in Eq.~\eqref{eq:FSDMDipInterDominated}. In this case $\mathbbm{W}^{\mathrm{dip},+}$ and $\mathbbm{W}^{\mathrm{int},+},$ have the structure

\smallskip{}
\begin{equation}
\mathbb{W}^{\mathrm{dip},+}=
\begin{bmatrix}
\mathbbm{N}_{\mathrm{XX}}^{\mathrm{dip},+}+\langle\mathbbm{N}_{\mathrm{XX}}^{\mathrm{dip},+}\rangle & \,\mathbbm{N}_{\mathrm{XY}}^{\mathrm{dip},+}-\langle\mathbbm{N}_{\mathrm{XY}}^{\mathrm{dip},+}\rangle  \\[4pt]
\,\mathbbm{N}_{\mathrm{XY}}^{\mathrm{dip},+}-\langle\mathbbm{N}_{\mathrm{XY}}^{\mathrm{dip},+}\rangle & -\left(\mathbbm{N}_{\mathrm{XX}}^{\mathrm{dip},+}+\langle\mathbbm{N}_{\mathrm{XX}}^{\mathrm{dip},+}\rangle \right)\end{bmatrix},
\label{eq:WdipintDip}
\end{equation}

and

\smallskip{}
\begin{equation}
\mathbb{W}^{\mathrm{int},+}=
\begin{bmatrix}
0& \,\mathbbm{N}_{\mathrm{XY}}^{\mathrm{int},+}-\langle\mathbbm{N}_{\mathrm{XY}}^{\mathrm{int},+}\rangle  \\[4pt]
\,-\left(\mathbbm{N}_{\mathrm{XY}}^{\mathrm{int},+}-\langle\mathbbm{N}_{\mathrm{XY}}^{\mathrm{int},+}\rangle \right) & 0
\end{bmatrix},
\label{eq:WdipintInt}
\end{equation}

\noindent where $\mathbbm{N}_{\mathrm{XX}}^{\mathrm{dip},+}+\langle\mathbbm{N}_{\mathrm{XX}}^{\mathrm{dip},+}\rangle$ is a real matrix,  $\mathbbm{N}_{\mathrm{XY}}^{\mathrm{dip},+}-\langle\mathbbm{N}_{\mathrm{XY}}^{\mathrm{dip},+}\rangle$ is an imaginary matrix, and $\mathbbm{N}_{\mathrm{XY}}^{\mathrm{int},+}-\langle\mathbbm{N}_{\mathrm{XY}}^{\mathrm{int},+}\rangle$ is also imaginary, as well as, a tridiagonal matrix. Taking into consideration that the interlayer exchange interaction strength is two to three orders of magnitude larger than the dipolar interaction strength and that $\mathbbm{W}^{\mathrm{int},+}$ is a diagonalizable matrix, thus in virtue of the \textit{Bauer-Fike theorem}\cite{bauer1960norms}, the eigenvalues of the interlayer exchange interaction $\mathbbm{W}^{\mathrm{int},+}$ dominates the frequency shift, whereas the dipolar contribution $\mathbbm{W}^{\mathrm{dip},+}$ plays the role of inducing perturbative corrections. In this sense,  the frequency shift scale is governed by the interlayer interaction matrix $\mathbbm{P}^{\mathrm{int}}$ in $\mathbbm{W}^{\mathrm{int},+}$. In other words, the frequency shift will be dominated by the interlayer exchange interaction. 

We can reach the same conclusion in the general case where the magnitudes and norms of the counterpropagating modes at a given wave vector differ. In this case, the FSDM $\mathbbm{W}^{+}$ contains contributions from all interactions which can be written explicitly as

\begin{widetext}
\begin{equation}
\label{dispersion_text_smallkz}
\mathbb{W}^{\mathrm{dip},+}=
\begin{bmatrix}
\left(\mathbbm{N}_{\mathrm{XX}}^{\mathrm{dip},+}-\mathbbm{A}_{\mathrm{XX}}^{\mathrm{dip}}\right)-A_{\mathrm{XY}}^{\mathrm{dip}} & \,\mathbbm{N}_{\mathrm{XY}}^{\mathrm{dip},+}-\mathbbm{P}_{\mathrm{XY}}^{\mathrm{dip}}-\mathbbm{P}_{\mathrm{XX}}^{\mathrm{dip}}-\mathrm{i}\mathbbm{w}\, \mathbbm{R}_{12}\mathbbm{R}_{12} \\[4pt]
\,\mathbbm{N}_{\mathrm{XY}}^{\mathrm{dip},+}-\mathbbm{P}_{\mathrm{XY}}^{\mathrm{dip}}+\mathbbm{P}_{\mathrm{XX}}^{\mathrm{dip}}-\mathrm{i}\,\mathbbm{w}\left(\mathbbm{1}-\mathbbm{R}_{11}^{\dagger}\mathbbm{R}_{11}^{\dagger}\right) & -\left(\mathbbm{N}_{\mathrm{XX}}^{\mathrm{dip},+}-\mathbbm{A}_{\mathrm{XX}}^{\mathrm{dip}}\right)-A_{\mathrm{XY}}^{\mathrm{dip}}
\end{bmatrix},
\end{equation}
\begin{equation}
\label{dispersion_text_smallkz}
\mathbb{W}^{\mathrm{a},+}=
\begin{bmatrix}
\mathbbm{A}_{\mathrm{XY}}^{\mathrm{a}} & \,\mathbbm{N}_{\mathrm{XY}}^{\mathrm{a},+}-\mathbbm{P}_{\mathrm{XY}}^{\mathrm{a}} \\[4pt]
\,-\left(\mathbbm{N}_{\mathrm{XY}}^{\mathrm{a},+}-\mathbbm{P}_{\mathrm{XY}}^{\mathrm{a}}\right) & -\mathbbm{A}_{\mathrm{XY}}^{\mathrm{a}}
\end{bmatrix},
\end{equation}
\end{widetext}

\noindent where $\mathrm{a}\in\{\mathrm{int, ex, U, Ze}\}$. Here, the diagonal (off-diagonal) blocks of $\mathbbm{W}^{\mathrm{dip},+}$ are real (imaginary) matrices as in the previous case. In a similar fashion, when $\mathrm{a}=\mathrm{int}$ the diagonal (off-diagonal) blocks of $\mathbbm{W}^{\mathrm{int},+}$ are real (imaginary) and tridiagonal matrices. In the case of $\mathrm{a}\in\{\mathrm{ex, U, Ze}\}$, the diagonal (off-diagonal) blocks of $\mathbbm{W}^{\mathrm{a},+}$ are also real (imaginary) but diagonal matrices. Applying the same arguments than in the previous case, i.e., that the interlayer exchange interaction strength is two to three orders of magnitude larger than the dipolar, exchange, uniaxial anisotropy and Zemman interaction strengths, and that the  $\mathbbm{W}^{\mathrm{int},+}$ is diagonizable, thus the dipolar, exchange, uniaxial anisotropy and Zemman contribution ($\mathbbm{W}^{\mathrm{dip},+}$, $\mathbbm{W}^{\mathrm{ex},+}$, $\mathbbm{W}^{\mathrm{U},+}$, and $\mathbbm{W}^{\mathrm{Ze},+}$, respectively), induces perturbative corrections to the dominant interlayer exchange interaction eigenvalues \cite{bauer1960norms}, leaving the leading frequency shift scale governed by the interlayer interaction matrix  $\mathbbm{W}^{\mathrm{int},+}$ as well. 

A frequency shift dominated by the interlayer exchange interaction is consistent with the non-identical counterpropagating modes $\tilde{\mathbbm{m}}^{\pm}$ at the same wave-vector magnitude $k$ in the Damon-Eshbach configuration. The interlayer exchange interaction and other interactions contributions to their frequencies become unequal. However, because the interlayer exchange interaction typically exceeds other interaction strengths including dipolar one, the frequency shift resulting from the difference in spin wave frequencies is dominated by the interlayer exchange interaction. In contrast, in the special case where the mode profiles are identical, both counterpropagating modes possess the same degree of inhomogeneity. As a result, their interlayer (and intralayer) exchange contributions are equal, and the nonzero frequency shift arises exclusively from dipolar interactions. 

This analysis clarifies the physical picture in which dipolar interactions primarily drive mode asymmetry, while interlayer exchange interactions convert this asymmetry into a dominant frequency imbalance, thereby governing the magnitude of the frequency shift of non-reciprocal spin waves.

\subsection{Eccentricity $\epsilon_n$, dephase angle $\varphi_n^0$ and tilting angle $\tau_n$, and Rotational matrices $\mathbbm{R}_{11}$ and $\mathbbm{R}_{12}$}
The $N\times N$ diagonal matrices $\mathbbm{R}_{11}$ and $\mathbbm{R}_{12}$ play a fundamental role in the FSDM $\mathbbm{W}^{\pm}$, given that they weight the influence of each interaction in the frequency shift of the nonreciprocal dispersion relation, as mentioned in previous sections. As seen in Eqs. \eqref{eq:TNXXdip} - \eqref{eq:TNYXa}, these matrices weight the  dynamic matrix blocks $\mathbbm{N}_{\mathrm{XX}}^{\mathrm{dip}}$, $\mathbbm{N}_{\mathrm{XY}}^{\mathrm{a}}$ at the FSDM blocks, where $\mathrm{a}\in\{\mathrm{dip, int, ex, U, Ze}\}$. Accordingly, these matrices deserve an analysis. 

As said in previous sections, the n'th diagonal entries of $\mathbbm{R}_{11}$ and $\mathbbm{R}_{12}$, denoted as $R_{11}^n$ and $R_{12}^n$ respectively, correspond to the first line entries of the n'th unitary $2\times 2$ rotational matrix $\mathbbm{R}_n$ as seen in Eq. \eqref{eq:Riequation}, with $n=1,2,\cdots N$. Since $R_{11}^n$ and $R_{12}^n$ can be written in terms of the Eccentricity $\epsilon_n$, tilting angle $\varphi_n^0$ and phase shift $\tau_n$, as seen in Eqs. \eqref{eq:ReR11}-\eqref{eq:ImR12}, in the following we analyze the behaviour of $R_{11}^n$ and $R_{12}^n$ as function of the wavevector for all sublayers, correlating them with the orbit paramenters. To this end, let's analyze the orbit parameters of the fundamental SW mode as function of the wavevector for a NiFe homogeneous film with homogeneous magnetization, and the nonreciprocal MD and the GML systems, to analyze later the matrices $\mathbbm{R}_{11}^n$, and $\mathbbm{R}_{12}^n$.

\begin{figure*}[th!]
    \centering
    \includegraphics[scale=0.43]{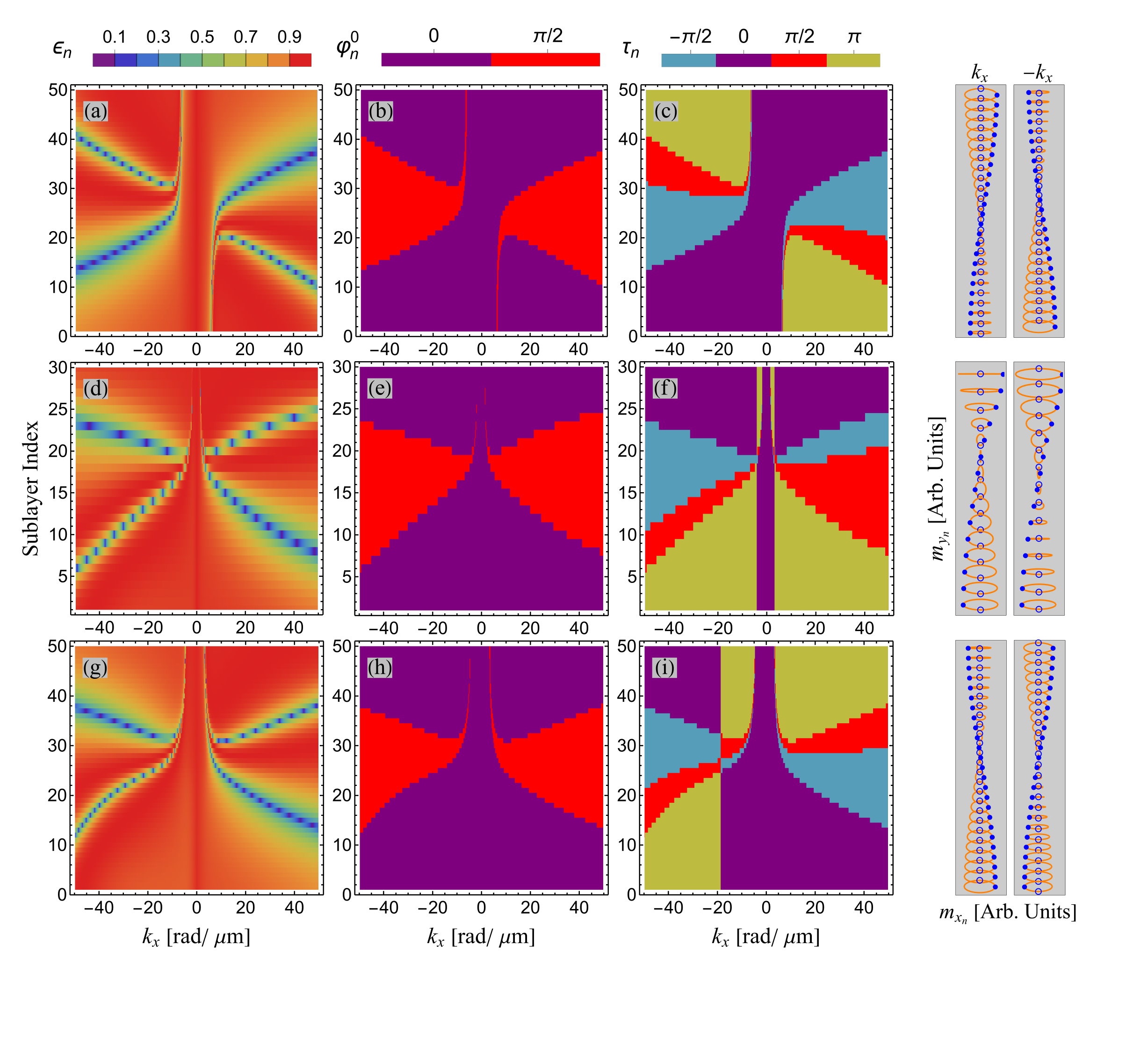}
    \caption{Spin wave orbit parameters distribution along the system thickness as function of the wavevector $k_x$. Eccentricity $\epsilon_n$, tilting angle $\varphi_n^0$ and phase shift angle $\tau_n$ are shown for three different multilayered systems with in-plane homogeneous magnetization parallel to $\hat z$: (a)-(c) a single layer of NiFe with saturation magnetization, stiffness constant, thickness and uniaxial anisotropy given as $M_s^{\text{NiFe}}=845 $ kA/m, $A_{\text{NiFe}}=12.8$ pJ/m, $d_{t,NiFe}=50$ nm, and $K_{\text{NiFe}}=0$, respectively; (d)-(f) a planar NiFe layer with graded magnetization saturation along the thickness whose profile changes from $M_s$ = 800 $\text{kA/m}$ to $M_s$ = 1600 $\text{kA/m}$, a thickness $d_{t,\text{NiFe}}=60$ nm and a stiffness constant $A_{\text{NiFe}}$ = 11 $\text{pJ/m}$. For this system, the applied field is $\mu_0 H=1.5 $ mT along the saturation magnetization direction; (g)-(i) a CoFeB/NiFe bilayer with saturation magnetization, stiffness constant, thickness and uniaxial anisotropy constant as $M_s^{\text{CoFeB}}$ = 1270 $\text{kA/m}$ ($M_s^{\text{NiFe}}$ = 845 $\text{kA/m}$), $A_{\text{CoFeB}}$ = 17 $\text{pJ/m}$ ($A_{\text{NiFe}}$ = 12.8 $\text{pJ/m}$), $d_{t,\text{CoFeB}}=25$ nm ($d_{t,\text{NiFe}}=25$ nm) and $K_{\text{CoFeB}}=0$ ($K_{\text{NiFe}}=0$), respectively, with an applied magnetic field $\mu_0H=30$ mT.}
    \label{fig:Excfasetaoorbi}
\end{figure*}

\begin{figure}[]
    \centering
   \includegraphics[scale=0.46]{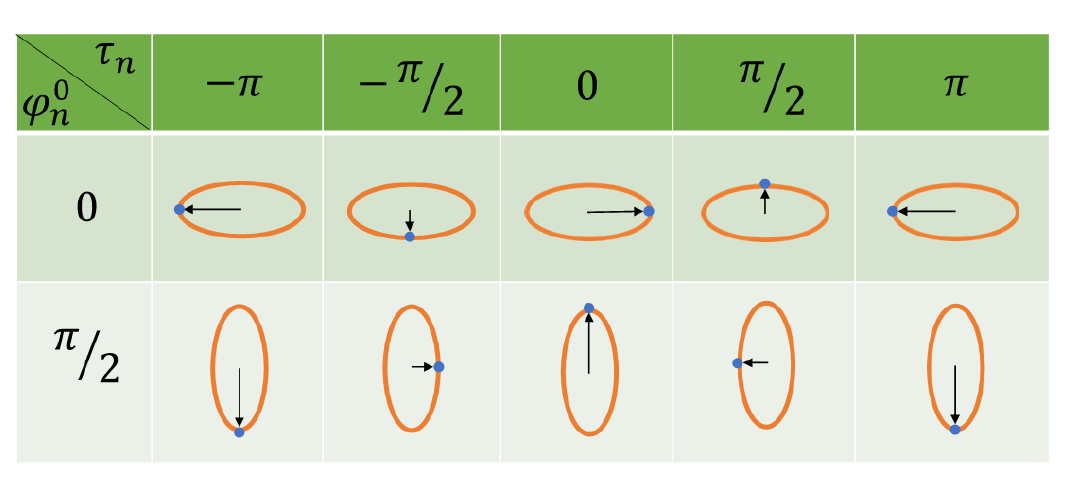} 
    \caption{A single layer of NiFe with in-plane homogeneous magnetization, and saturation magnetization,  stiffness constant, thickness and uniaxial anisotropy constant given as $M_s^{\text{NiFe}}$ = 845 $\text{kA/m}$, $A_{\text{NiFe}}$ = 12.8 $\text{pJ/m}$, $t_{\text{NiFe}}=50$ nm, and $K_{\text{NiFe}}=0$, respectively. }
    \label{fig:phaeorbit}
\end{figure}

The eccentricity $\epsilon_n$, phase shift $\tau_n$ and tilting angle $\varphi_n^0$ are calculated by obtaining the eigenmodes $\tilde{\mathbbm{m}}$ at a given wavevector $\mathbf{k}$, thus replacing them in Eqs. \eqref{eq:eccentricity}-\eqref{eq:phaseshift}. Fig. \ref{fig:Excfasetaoorbi} shows the orbit parameters of the fundamental mode as function of the wavevector and along each sublayer of the reciprocal NiFe layer (Fig. \ref{fig:Excfasetaoorbi}(a)-(c)), and the nonreciprocal GML (Fig. \ref{fig:Excfasetaoorbi}(d)-(f)) and MD (Fig. \ref{fig:Excfasetaoorbi}(g)-(i)) systems. Notice that the tilting angle has only two possible values $\varphi_n^0=0,\pi/2$ and the transition between them coincides with the boundary of zero eccentricity. In all our results, zero eccentricity defines the transition point where the larger and smaller semiaxis permute between them, which is therefore consistent with boundary in the tilting angle $\varphi_n^0$ that separates the region with $\varphi_n=0$ (larger semiaxis along $\hat{x}_n$) from the region with $\varphi_n^0=\pi/2$ (larger semiaxis along $\hat{y}_n$). One can also observe that the phase shift angle takes only discrete angles $\tau_n=\pm l\pi/2$ with $l=0,1,2,...$, with well-defined boundaries between them. The addition $\tilde{{\tau}}_n=\varphi_n^0+\tau_n$ defines the orientation of the dynamic magnetization at zero time $\mathbf{m}_{n}[t=0]$. Fig. \ref{fig:phaeorbit} shows the possible orientations of the $\mathbf{m}_{n}[t=0]$ according to the addition of the possible values of $\varphi_n^{n}$ and $\tau_n$, and is very useful to understand the SW orbit distribution along the thickness shown at the two rightmost column of Fig. \ref{fig:Excfasetaoorbi} at $k_x=\pm 20$ rads$/\mu$m. Here, the blue-circle and empty blue-circle points correspond to $\mathbf{m}_{n}[t=0]$ and $\mathbf{m}_{n}[t=\pi/2]$, respectively. Notice the inversion symmetry of the orbit parameters (i.e., $\epsilon_n[k_x]=\epsilon_{N-n}[-k_x]$, $\varphi^0_n[k_x]=\varphi^0_{N-n}[-k_x]$, and $\tau_n[k_x]=\tau_{N-n}[-k_x]$) which is consistent with the SW mode heterosymmetry of the  reciprocal NiFe homogeneous layer (see Figs. \ref{fig:Excfasetaoorbi}(a)-(c)). In contrast, the lack of inversion symmetry of the orbit parameters in the nonreciprocal MD and GML systems (as shown in Figs. \ref{fig:Excfasetaoorbi}(d)-(h)) is consistent with that the counterpropagating SW modes at the same wavevector magnitude $k_x$ are no more heterosymmetric. According to this, one could say that Fig. \ref{fig:Excfasetaoorbi} gives a widespread overview of the mode inhomogeneity at a given range of wavevectors, therefore, useful to get an idea on how nonreciprocal the system is. 

\begin{figure}[th!]
    \centering
    \includegraphics[scale=0.073]{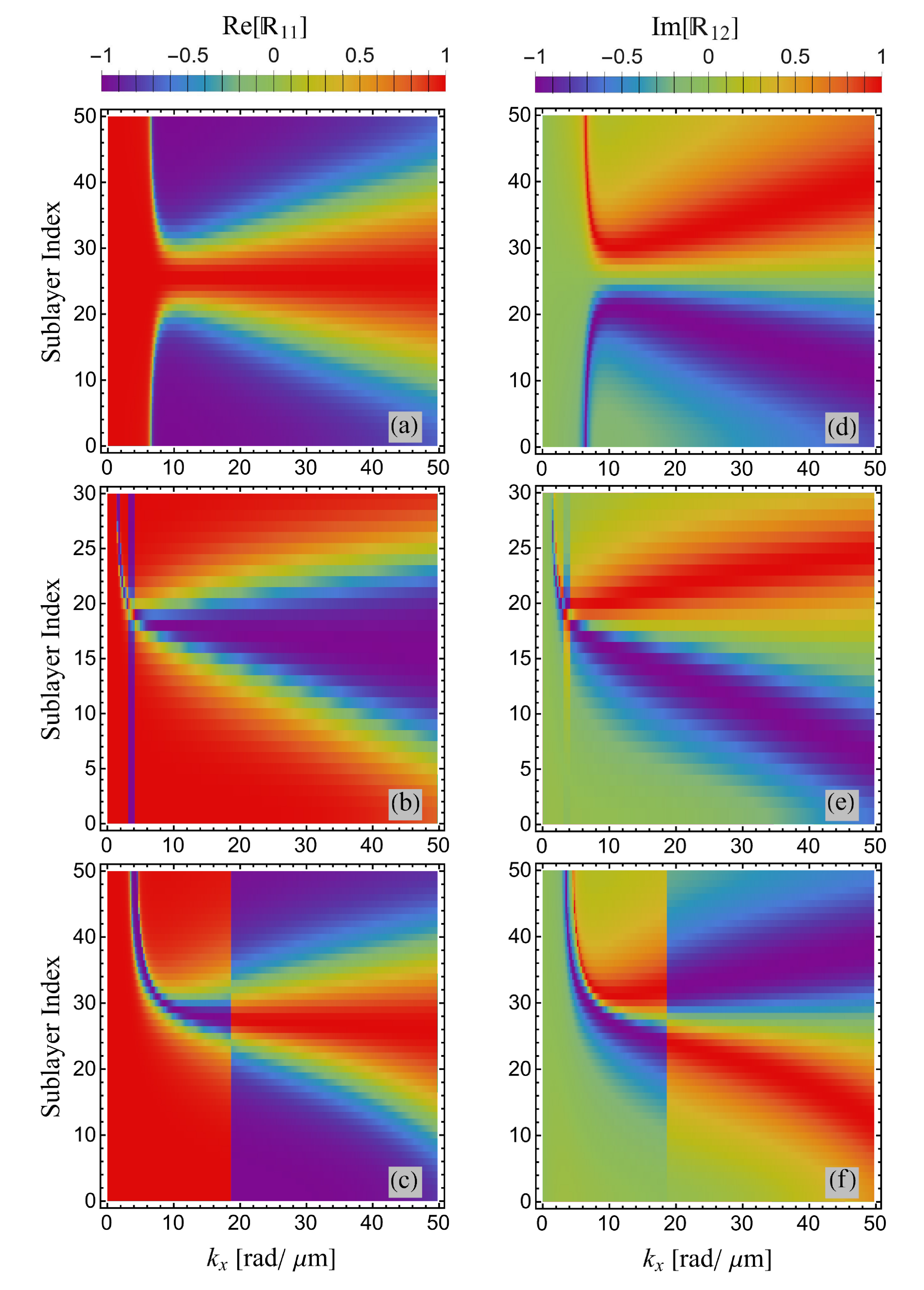} 
    \caption{$N\times N$ diagonal matrices profile along the system thickness and as function of the wavevector magnitude $k_x$ for the fundamental SW order mode in a Damon-Esbach configuration and homogeneous magnetization. (a)-(c) and (d)-(f) show the profile of $\mathbbm{R}_{11}=\text{Re}[\mathbbm{R}_{11}]$ and $\mathbbm{R}_{12}=\text{Im}[\mathbbm{R}_{12}]$, respectively. Matrices comparison among three systems: (a) and (d) for a reciprocal and homogeneous single NiFe layer, with saturation magnetization, stiffness constant, thickness and uniaxial anisotropy given as $M_s^{\text{NiFe}}=845 $ kA/m, $A_{\text{NiFe}}=12.8$ pJ/m, $d_{t,NiFe}=50$ nm, and $K_{\text{NiFe}}=0$, respectively; (b) and (e) corresponding to a planar NiFe layer with graded magnetization saturation along the thickness whose profile changes from $M_s$ = 800 $\text{kA/m}$ to $M_s$ = 1600 $\text{kA/m}$, a thickness $d_{t,\text{NiFe}}=60$ nm and a stiffness constant $A_{\text{NiFe}}$ = 11 $\text{pJ/m}$. For this system, the applied field is $\mu_0 H=1.5 $ mT along the saturation magnetization direction; and (c) and (f) corresponding to a CoFeB/NiFe bilayer with saturation magnetization, stiffness constant, thickness and uniaxial anisotropy constant as $M_s^{\text{CoFeB}}$ = 1270 $\text{kA/m}$ ($M_s^{\text{NiFe}}$ = 845 $\text{kA/m}$), $A_{\text{CoFeB}}$ = 17 $\text{pJ/m}$ ($A_{\text{NiFe}}$ = 12.8 $\text{pJ/m}$), $d_{t,\text{CoFeB}}=25$ nm ($d_{t,\text{NiFe}}=25$ nm) and $K_{\text{CoFeB}}=0$ ($K_{\text{NiFe}}=0$), respectively, with an applied magnetic field $\mu_0H=30$ mT.}
    \label{fig:ReImR11R12}
\end{figure}

According to Eqs. \eqref{eq:ReR11}-\eqref{eq:ImR12} and noticing that the orbit parameters differences between two counterprogating SW modes at the same wavevector magnitude $k_x$ are such that $\Delta \phi_n^0=0,\pm\pi/2,\pm \pi$ and $\Delta \tau_n=\pm l\pi/2$ with $l=0,1,2,\cdots$, thus the entries of the $N\times N$ diagonal matrices $\mathbbm{R}_{11}$ and $\mathbbm{R}_{12}$ are real or  imaginary numbers. As mentioned in previous sections,  $\mathbbm{R}_{11}$ and $\mathbbm{R}_{12}$ are mutually exclusive real or  imaginary matrices (i.e., if $\mathbbm{R}_{11}$ is real (imaginary), thus $\mathbbm{R}_{12}$ is imaginary (real)). At the fundamental counterpropagating order mode, $\mathbbm{R}_{11}$ and $\mathbbm{R}_{12}$ turns out to be real and  pure imaginary matrices, respectively. 
In Fig. \ref{fig:ReImR11R12}, we show these matrices for the fundamental order mode of the reciprocal homogeneous NiFe layer with homogeneous magnetization pointing along $\hat{z}$ (see Fig. \ref{fig:ReImR11R12}(a) and (d)), the MD system (see Fig. \ref{fig:ReImR11R12}(b) and (e)) and GML system (see Fig. \ref{fig:ReImR11R12}(c) and (f)).

In the case of the reciprocal homogeneous NiFe layer with homogeneous magnetization, the inversion symmetry of the orbit parameters is consistent with the even symmetry $R_{11}^n=R_{11}^{N-n}$ and the odd symmetry  $R_{12}^n=-R_{12}^{N-n}$ along the thickness of the magnetic layer, as seen in Fig. \ref{fig:ReImR11R12}(a), (d), (g), and (j) for the fundamental SW order mode. The same symmetries apply to higher SW order modes in the reciprocal layer. These results helps to contrast with the nonreciprocal MD and GML systems, featured by the lack of inversion symmetry in their orbit parameters, which is consistent with the lack of even or odd symmetries in the same matrix entries. This lack of symmetry can be considered as a signature of a nonreciprocal dispersion relation dominated mostly by the interlayer exchange interaction.     


\section{Conclusions}

We have presented a unified theoretical framework to identify the microscopic origin of spin-wave nonreciprocal frequency shift in planar multilayer magnetic heterostructures without Dzyaloshinskii-Moriya interaction. Using a frequency-shift dynamic matrix, we showed that the dispersion asymmetry generally cannot be attributed solely to dipolar interactions. Instead, once counter-propagating modes differ in their geometric structure along the thickness - a generic situation in multilayer systems beyond the thin-film limit - the frequency shift is dominated by exchange interactions, with the regular isotropic exchange along the film thickness (referred to as interlayer exchange throughout the manuscript) providing the leading contribution.

Purely dipolar nonreciprocal frequency shift emerges only in the special case where the normalized spin-wave profiles at $\pm\mathbf{k}$ are identical layer by layer. In all other cases, the dominant contribution to the frequency shift originates from the imbalance of exchange-mediated energy by counter-propagating modes. By evaluating realistic parameter ranges and analyzing two representative systems - a graded-magnetization NiFe film and a CoFeB/NiFe bilayer - we demonstrated that the interlayer exchange contribution exceeds the dipolar, intralayer exchange, uniaxial and Zeeman contributions by up to two to three orders of magnitude over a broad wave-vector range.

Finally, we examined the symmetry properties of spin-wave orbits through their eccentricity $\epsilon_n$, tilt angle $\varphi_n^0$, and phase shift $\tau_n$. For a homogeneous NiFe layer with reciprocal dispersion, these orbit parameters exhibit inversion symmetry, consistent with the heterosymmetric nature of the counterpropagating modes. In contrast, both the magnonic diode and graded-magnetization systems display nonreciprocal dispersion and nonheterosymmetric mode profiles, leading to broken inversion symmetry in the orbit parameters. These symmetry differences are directly reflected in the structure of the matrices $\mathbbm{R}_{11}$ and $\mathbbm{R}_{12}$, which exhibit even and odd thickness symmetries, respectively, in the homogeneous case, and lack such symmetries in the nonreciprocal systems.

Physically, dipolar interactions generate asymmetric mode profiles, while the energetic cost of this asymmetry is set primarily by interlayer exchange, which strongly penalizes profile inhomogeneity. Expressing the frequency-shift dynamic matrix in terms of orbit geometry parameters clarifies how these geometric differences are converted into large frequency shifts. Our results revise the common interpretation of nonreciprocal spin-wave dispersion in multilayers and establish isotropic exchange as the leading interaction responsible for the frequency shift. With this, we provide fundamental physical insight with direct implications for the design and optimization of multilayer magnonic devices, such as diodes, isolators, and directional wave-based computing architectures, which might encourage further developments in the emerging field of 3D magnon devices for improving the capabilities of transferring, buffering and processing information.

\section*{Declaration of competing interest}
The authors declare that they have no known competing financial interests or personal relationships that could have appeared to influence the work reported in this paper.

\begin{acknowledgments}
This research has received funding support from Chilean Doctorado Nacional ANID via fellowship Grant 21211429.
\end{acknowledgments}

\section*{Data Availability Statement}

The data are available upon reasonable request from the authors.

\section*{CRediT authorship contribution statement}
 \textbf{Claudia Negrete}: Writing - original draft, Visualization, Conceptualization, Investigation, Formal Analysis. \textbf{Attila K\'akay}: Writing - review \&
editing, Conceptualization, Investigation. \textbf{Jorge A. Ot\'alora}: Writing - review \& editing, Conceptualization, Supervision, Resources, Project administration, Methodology, Investigation, Formal Analysis.

\smallskip{}

\bibliographystyle{apsrev4-2} 
\bibliography{References}
\end{document}